\documentclass[journal]{IEEEtran}
\IEEEoverridecommandlockouts
\usepackage{cite}
\usepackage{amsmath,amssymb,amsfonts}
\usepackage{algpseudocodex}
\usepackage{algorithm}
\algnewcommand{\LeftComment}[1]{\Statex \(\triangleright\) #1}
\usepackage{graphicx}
\usepackage{textcomp}
\usepackage{xcolor}
\usepackage{url}
\usepackage{stackengine}
\usepackage{multirow}
\def\BibTeX{{\rm B\kern-.05em{\sc i\kern-.025em b}\kern-.08em
    T\kern-.1667em\lower.7ex\hbox{E}\kern-.125emX}}
\begin{document}

\title{SpikePipe: Accelerated Training of Spiking Neural Networks via Inter-Layer Pipelining and Multiprocessor Scheduling}

\author{\IEEEauthorblockN{Sai Sanjeet\IEEEauthorrefmark{2}, Bibhu Datta Sahoo\IEEEauthorrefmark{2}, and Keshab K. Parhi, {\em Fellow, IEEE}\IEEEauthorrefmark{3}}\\
\IEEEauthorblockA{\IEEEauthorrefmark{2}Dept. of Electrical Eng., University at Buffalo, Buffalo, NY, USA. e-mail: bibhu@buffalo.edu}\\
\IEEEauthorblockA{\IEEEauthorrefmark{3}Dept. of Electrical and Computer Eng., University of Minnesota, Minneapolis, USA. e-mail: parhi@umn.edu} \vspace{-2em}
\thanks{This work was supported in part by the National Science Foundation under grant number CCF-1954749..}
}

\maketitle
\begin{abstract}
Spiking Neural Networks (SNNs) have gained popularity due to their high energy efficiency. Prior works have proposed various methods for training SNNs, including backpropagation-based methods. Training SNNs is computationally expensive compared to their conventional counterparts and would benefit from multiprocessor hardware acceleration. This is the first paper to propose inter-layer pipelining to accelerate training in SNNs using systolic array-based processors and multiprocessor scheduling. The impact of training using delayed gradients is observed using three networks training on different datasets, showing no degradation for small networks and $<10\%$ degradation for large networks. The mapping of various training tasks of the SNN onto systolic arrays is formulated, and the proposed scheduling method is evaluated on the three networks. The results are compared against standard pipelining algorithms. The results show that the proposed method achieves an average speedup of 1.6$\times$ compared to standard pipelining algorithms, with an upwards of 2$\times$ improvement in some cases. The incurred communication overhead due to the proposed method is less than $0.5\%$ of the total required communication of training.
\end{abstract}
\begin{IEEEkeywords}
Spiking neural networks, pipelining, multiprocessor scheduling, hardware accelerators.
\end{IEEEkeywords} \vspace{-1em}
\section{Introduction}
\indent Spiking neural networks (SNNs) are a type of neural network that mimic the functionality of biological neural networks \cite{Maass1997, Gerstner2002, Pfeiffer2018, Parhi2020}. Due to their high energy efficiency, they have been employed in a wide range of applications, including computer vision, robotics, and speech recognition, and can achieve performance similar to conventional neural networks \cite{Tavanaei2019}. SNNs can be divided into two main types: synchronous \cite{Yujie2018,Fang2020,Anwani2020} and asynchronous \cite{Pfeiffer2016,Stromatias2017,Thiele2018}. In synchronous SNNs, neuronal outputs are computed at linearly-spaced time intervals, while asynchronous or event-driven SNNs compute neuronal outputs based on the arrival of spikes.

Both synchronous and asynchronous SNNs use models of neurons to compute membrane potentials and spike timings. There are various models of neurons, such as the integrate-and-fire (I\&F) \cite{LIF1999}, the leaky integrate-and-fire (LIF) \cite{LIF1999}, the Izhevsky model \cite{Izhikevich2003}, and the Hodgkin-Huxley model \cite{Hodgkin1952}. The simplest model, the I\&F model, is a linear model that does not account for the time-dependent nature of the membrane potential. The LIF model is a nonlinear model that incorporates decay and accounts for the time-dependent nature of the membrane potential. The Hodgkin-Huxley model, the most accurate model of the neuron, accounts for the time-dependent nature of the membrane potential and the ion channels in the neuron. However, simulating the Hodgkin-Huxley model is computationally expensive.

In the most straightforward implementation, the LIF neuron is modeled as a first-order infinite impulse response (IIR) filter \cite{Sahoo2017, Sanjeet2023, Fang2020}, where the IIR filter's output represents the neuron's membrane potential. Once the membrane potential reaches a threshold value, the neuron spikes, and the membrane potential is reset to resting potential. In a digital implementation, the IIR filter is implemented using a digital first-order section.

There are various methods of training SNNs, with the two most common being backpropagation \cite{Pfeiffer2016, Yujie2018, Fang2020, Stromatias2017, Anwani2020, Lee2020} and Spike Timing Dependent Plasticity (STDP) \cite{Dan1992, Lee2018, Thiele2018, Lee2019}. Backpropagation, which inherently introduces feedback loops, is challenging to map onto multiple processors to achieve high throughput.

Prior works have proposed various methods for training conventional neural networks, such as Convolutional Neural Networks (CNNs), on multiple processors \cite{Bennun2019, Huang2019, Narayanan2019, Zhao2020, Wang2020, Unnikrishnan2021}. In our prior LayerPipe approach \cite{Unnikrishnan2021}, we proposed the use of variable delayed gradients to achieve inter-layer pipelining of CNNs. Delayed gradients have been used to pipeline adaptive digital filters \cite{Long1989}. The basic assumption is that the gradients can be replaced by delayed gradients if the gradients are slowly varying. However, these methods cannot be directly applied to SNNs because they do not account for the fact that activations in SNNs are binary. By taking this into account, it is possible to further pipeline the training process of training the SNNs at a fine-grain level without significant overhead in communication. It may be noted that there exist prior works that focused on developing hardware for accelerating the training of SNNs \cite{Liang2022,Singh2022,Yin2023}. However, this work's primary objective is accelerating training by efficient mapping onto multiple processors, and therefore, can be used to further accelerate the prior proposed hardware with small modifications.

The contributions of this paper are three-fold. We consider training a digital SNN model that is based on a first-order IIR digital filter with a forward gain that is not unity, introduced in our prior work \cite{Sahoo2017, Sanjeet2023}. Note that in prior SNNs, the forward gain of the first-order IIR filter had been considered to be unity. Simulation results show that the proposed architecture achieves $98.6\%$ accuracy on the MNIST dataset using the proposed structure, compared to $98.3\%$ with a structure with unity gain in the forward path. The hardware overhead for realizing non-unity gain is also minimal. Our first contribution lies in deriving the backpropagation equations for training the proposed SNN architecture. It is shown that the IIR filter structure used in the forward pass of the neuron model can also be used to compute the backward pass, implying no additional specialized blocks are necessary for training the SNN. This is non-intuitive, and is important because the same datapath can be used for both inference and training, especially in edge devices. Second, we exploit a variable delayed-gradient approach to achieve inter-layer pipelining. This is inspired by our prior work on accelerating training in CNNs using Layerpipe \cite{Unnikrishnan2021}. This paper is the first to accelerate training in SNNs using inter-layer pipelining. This creates concurrency that can be exploited to map the computations of different layers to multiple processors. It is shown that the use of delayed gradients does not degrade the accuracy by a significant amount. Third, a fine-grained layer-wise scheduling algorithm is proposed to map the tasks to multiprocessors to accelerate the training of SNNs. The proposed fine-grained pipelining algorithm is evaluated using a few sample networks, and the results are compared with existing scheduling algorithms.

This paper is organized as follows. Section II reviews the SNN model considered in this paper. Section III derives the backpropagation equations for training the SNN, and introduces the delayed gradient approach. Section IV formulates the mapping of the training tasks to systolic arrays. Section V presents a fine-grained scheduling algorithm to map the training tasks to multiple processors such that the underutilization of the processors is minimized. Simulation results are presented in Section VI.
\vspace{-1em}
\section{Modeling Neurons for SNNs} 
\begin{figure}[htbp]
    \centering
    \vspace{-0.5em}
    \includegraphics[scale=0.8]{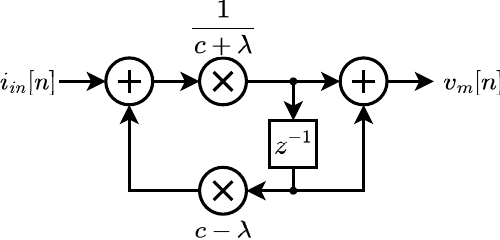} \vspace{-1em}
    \caption{Digital first-order section for implementing the LIF neuron model.} \vspace{-0.25em}
    \label{fig:iir_fos}
\end{figure}

\vspace{-0.5em}
\label{sec:modeling_neurons}
The first step in designing an SNN is modeling a single neuron. Of the various neuron models, the Leaky Integrate-and-Fire (LIF) \cite{LIF1999} model is used in this work. The LIF model is a nonlinear model that incorporates decay and accounts for the time-dependent nature of the membrane potential. The membrane potential, $v_m(t)$, of the LIF neuron follows first-order dynamics and is described by (\ref{eq:membrane_potential_lif}). The neuron produces a spike when the membrane potential crosses a threshold and is reset to a resting potential.
\begin{equation}
    \label{eq:membrane_potential_lif}
    C_m \frac{dv_m(t)}{dt} = i_{in}(t) - \frac{v_m(t) - V_{rest}}{R_m}
\end{equation}
\noindent where $i_{in}(t)$ is the net input current to the neuron, $C_m$ is the membrane capacitance, and $R_m$ is the membrane resistance that causes decay in the membrane potential. For simplicity, the resting potential $V_{rest}$ is taken to be zero. The Laplace transform of (\ref{eq:membrane_potential_lif}) is given by (\ref{eq:membrane_potential_lif_laplace}).
\begin{equation}
    \label{eq:membrane_potential_lif_laplace}
    V_m(s) = \frac{I_{in}(s)}{s \cdot C_m + \frac{1}{R_m}}
\end{equation}
\noindent where, $V_m(s)$ and $I_{in}(s)$ are the Laplace transforms of $v_m(t)$ and $i_{in}(t)$, respectively. From (\ref{eq:membrane_potential_lif_laplace}), it is evident that the transfer function from input currents to membrane potential is a first-order infinite impulse response (IIR) filter. The digital equivalent of this filter can be obtained using the bilinear transform as shown in (\ref{eq:membrane_potential_lif_digital}).
\begin{equation}
    \label{eq:membrane_potential_lif_digital}
    H(z) = \frac{V_m(z)}{I_{in}(z)} = \frac{1 + z^{-1}}{(c + \lambda) - (c - \lambda)z^{-1}}
\end{equation}
\noindent where, $c = 2C_m/T_s$, $T_s$ is the sampling period, and $\lambda = 1 / R_m$. The digital filter in (\ref{eq:membrane_potential_lif_digital}) is implemented using a digital first-order section as shown in Fig. \ref{fig:iir_fos}.

\begin{figure}[htbp]
    \centering
    \vspace{-1em}
    \includegraphics[scale=0.8]{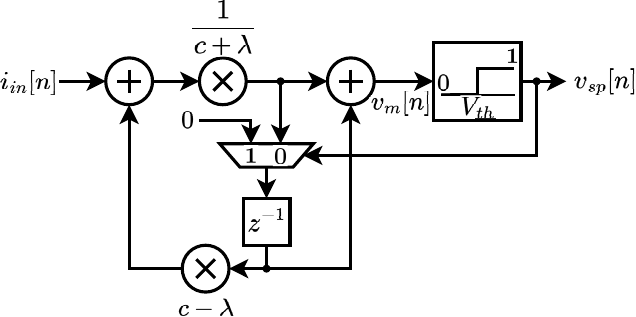} \vspace{-1em}
    \caption{Modified LIF structure incorporating the reset operation. The delay element is reset when the membrane potential crosses the threshold voltage.} \vspace{-1em}
    \label{fig:reset}
\end{figure}

The output of the first-order section is the membrane potential of the neuron. Once the membrane potential reaches a threshold value, the neuron spikes, and the membrane potential is reset to resting potential. The spiking operation is implemented using a simple comparator as shown in Fig. \ref{fig:reset}\footnote{It is to be noted that resetting the delay element also resets the feedforward path, making it harder to produce consecutive spikes. This introduces a soft refractory behavior in the neuron.}. $v_{sp}[n]$ is a binary time series whose value is 1 if there is a spike at timestep $n$ and 0 otherwise. The relation between the input current, the membrane potential, and the output spike train is given by (\ref{eq:difference_equation_lif}) and (\ref{eq:spike_train_lif}).
\begin{figure}[H]
    \centering
    \vspace{-1em}
    \includegraphics[scale=0.3]{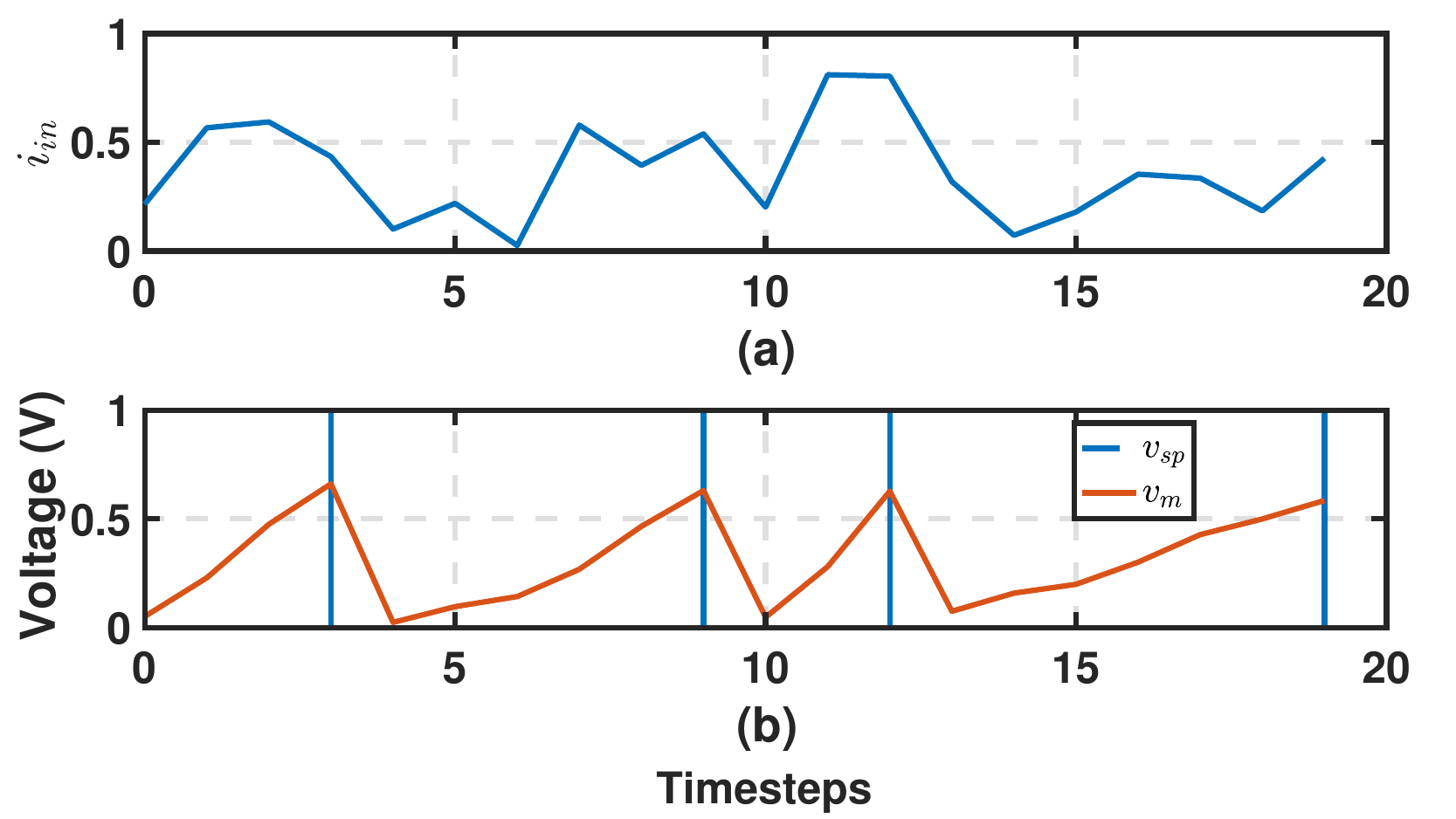} \vspace{-1em}
    \caption{(a) Sample input current for 20 timesteps and (b) the corresponding membrane potential and spike train. The values of $c$ and $\lambda$ are 4 and 0.25, respectively, and the threshold voltage is 0.5.} \vspace{-2em}
    \label{fig:sample_lif}
\end{figure}
\begin{align}
    \label{eq:difference_equation_lif}
    v_m[n] &= \frac{i_{in}[n] + (c - \lambda) \cdot d[n-1]}{c + \lambda} + d[n-1] \\
    d[n] &= \bar{v}_{sp}[n] \cdot \frac{i_{in}[n] + (c - \lambda) \cdot d[n-1]}{c + \lambda} \nonumber \\
    \label{eq:spike_train_lif}
    v_{sp}[n] &= \begin{cases}
        1, & \text{if } v_m[n] \geq V_{th} \\
        0, & \text{otherwise}
    \end{cases}
\end{align}
\noindent where, $d[n]$ is the intermediate value stored in the delay element at timestep $n$, and $\bar{v}_{sp}$ is the binary complement of $v_{sp}$. Figure \ref{fig:sample_lif} shows the membrane potential and the spike train for a sample input current.
\section{Training SNNs}

A neural network is a network of neurons interconnected by synapses. The topology of the network is generally predetermined and does not change over the course of training. In the proposed Spiking Neural Network (SNN), each neuron is modeled using the LIF neuron model as described in Section \ref{sec:modeling_neurons}. The neurons are arranged in layers, and synapses connect each neuron to a subset of neurons in the previous layer. The weights associated with the synapses are learned during the training process.

\begin{figure}[htbp]
    \centering
    \vspace{-1em}
    \includegraphics[scale=0.8]{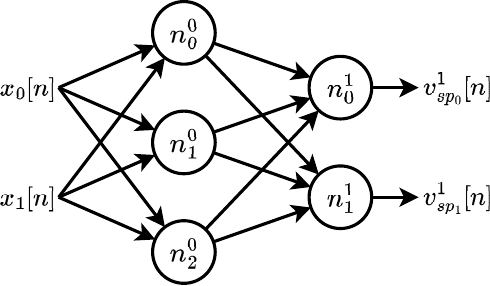} \vspace{-0.5em}
    \caption{A sample spiking neural network with two layers.} \vspace{-0.5em}
    \label{fig:sample_network}
\end{figure}

Figure \ref{fig:sample_network} shows an example network with two inputs, $x_0[n]$ and $x_1[n]$, one hidden layer with three neurons, and an output layer with two neurons. The neuron $i$ in layer $l$ is represented by $n_i^l$. The same naming convention is used for all intermediate variables $i_{in}$, $v_{m}$, and $v_{sp}$. The weight connecting input $i$ to neuron $j$ of the hidden layer is given by $w^0_{ij}$. Therefore, the input current to neuron $j$ of the hidden layer is given by (\ref{eq:input_current_hidden_layer}). \vspace{-0.5em}
\begin{equation}
    \label{eq:input_current_hidden_layer}
    i_{in_j}^0[n] = \sum_{i=0}^1 w^0_{ij} \cdot x_i[n]
\end{equation} \vspace{-1em}
\begin{figure}[bp]
    \centering
    \vspace{-1em}
    \includegraphics[scale=0.8]{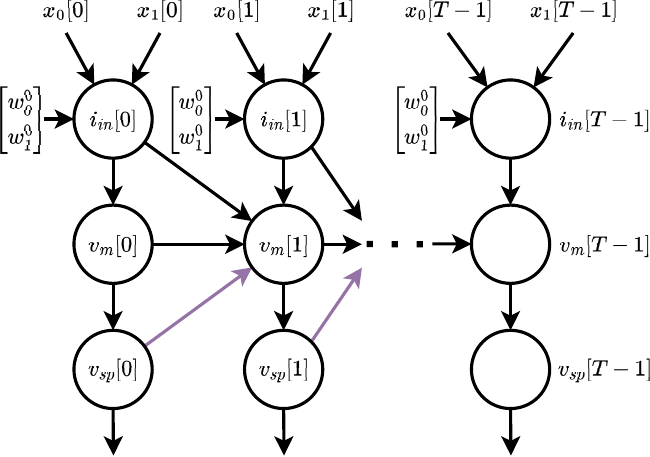} \vspace{-0.5em}
    \caption{Dataflow graph of a neuron in the hidden layer for $T$ timesteps. Spiking of the neuron in the previous timestep gates the input current and membrane potential of the current timestep, and is shown by the colored line.} \vspace{-1em}
    \label{fig:neuron_dfg}
\end{figure}

Similarly, the weight connected to neuron $i$ of the hidden layer and neuron $j$ of the output layer is given by $w^1_{ij}$. The corresponding input currents are used to compute the membrane potentials and spike trains of every neuron using (\ref{eq:difference_equation_lif}) and (\ref{eq:spike_train_lif}). The output of the network is the spike train of the output neurons. Figure \ref{fig:neuron_dfg} shows the unrolled dataflow graph (DFG) of a neuron in the hidden layer for $T$ timesteps, where nodes $i_{in}$, $v_m$, and $v_{sp}$ are computed using (\ref{eq:input_current_hidden_layer}), (\ref{eq:difference_equation_lif}), and (\ref{eq:spike_train_lif}), respectively.
\subsection{Loss Function}
For supervised learning, the network is trained using a loss function that measures the deviation between the output of the network and the desired output. The loss function is a function of the weights of the network, and the goal of the training process is to obtain the optimal set of weights that minimize the loss function. The use of the appropriate loss function is essential for training the network. Before discussing the loss function, it is crucial to understand the network's output. Spike trains are binary sequences, and encoding the desired output as a binary sequence is not trivial. Therefore, the output of the network is taken to be the membrane potential of the output neurons in the last timestep. To ensure that the membrane potential accurately reflects the network's output, the output neurons are transformed into accumulators with a feedforward path by removing their leaky nature and preventing the membrane potential from resetting, \emph{i.e.,} by setting $c=1$ and $\lambda=0$.

Using the membrane potential in the final timestep as the network output, the categorical cross-entropy loss function, given by (\ref{eq:loss_function}), is used.
\begin{equation}
    \label{eq:loss_function}
    \mathcal{L} = -\sum_{i=0}^{N-1} y_i \cdot \log\left( \frac{\exp(v_{m_i}^1[T-1])}{\sum_{j=0}^{N-1}\exp(v_{m_j}^1[T-1])} \right)
\end{equation}
\noindent where, $y = [y_0, y_1, \dots, y_{N-1}]^T$ is the one-hot encoded desired output, $N$ is the number of outputs of the network, $\exp(.)$ is the exponential function, and $\mathcal{L}$ is the loss between the network output and desired output. \vspace{-1em}
\subsection{Backpropagation}
To train the network, the gradient of the loss function with respect to the weights of the network is needed. This gradient is used to update the weights of the network using gradient descent. The gradient of the loss function with respect to the weights is computed using backpropagation. The backpropagation algorithm is based on the chain rule of differentiation. For each node in the DFG in Fig. \ref{fig:neuron_dfg}, the forward pass and backward pass equations are detailed. The forward pass equations are used to compute the output of the network, and the backward pass equations are used to calculate the gradient of the loss function with respect to the weights and intermediate outputs of the network.

\subsubsection{$i_{in}[n]$}

The forward pass equation for computing $i_{in}[n]$ of the first layer is given by (\ref{eq:forward_pass_iin}).
\begin{align}
    \label{eq:forward_pass_iin}
    i_{in_j}^0[n] &= \sum_{i} w^0_{ij} \cdot x_i[n]
\end{align}
For each subsequent layer, the forward pass equation for computing $i_{in}[n]$ is given by (\ref{eq:forward_pass_iin_subsequent}).
\begin{equation}
    \label{eq:forward_pass_iin_subsequent}
    i_{in_j}^l[n] = \sum_{i} w^l_{ij} \cdot v_{sp_i}^{l-1}[n]
\end{equation}
Generally, the forward pass equation for computing $i_{in}[n]$ is the same as for other neural networks such as CNNs. The only difference is the operation is repeated for every timestep. Therefore, in a Spiking CNN for each timestep, the input spike train is convolved with the weights to generate the input current to the neurons in the next layer.

The local gradients for the backward pass are given by (\ref{eq:backward_pass_local_iin_vsp}) and (\ref{eq:backward_pass_local_iin_wij}). \vspace{-1em}
\begin{align}
    \label{eq:backward_pass_local_iin_vsp}
    \frac{\partial i_{in_j}^l[n]}{\partial v_{sp_i}^{l-1}[n]} &= w^l_{ij} \\
    \label{eq:backward_pass_local_iin_wij}
    \frac{\partial i_{in_j}^l[n]}{\partial w^l_{ij}} &= v_{sp_i}^{l-1}[n]
\end{align}
The gradients with respect to the loss function are given by (\ref{eq:backward_pass_global_iin_vsp}) and (\ref{eq:backward_pass_global_iin_wij}) using the chain rule of differentiation.
\begin{align}
    \label{eq:backward_pass_global_iin_vsp}
    \frac{\partial \mathcal{L}}{\partial v_{sp_i}^{l-1}[n]} = \sum_j \frac{\partial \mathcal{L}}{\partial i_{in_j}^l[n]} \cdot \frac{\partial i_{in_j}^l[n]}{\partial v_{sp_i}^{l-1}[n]} = \sum_j \frac{\partial \mathcal{L}}{\partial i_{in_j}^l[n]} \cdot w^l_{ij} \\
    \label{eq:backward_pass_global_iin_wij}
    \frac{\partial \mathcal{L}}{\partial w^l_{ij}} = \sum_n \frac{\partial \mathcal{L}}{\partial i_{in_j}^l[n]} \cdot \frac{\partial i_{in_j}^l[n]}{\partial w^l_{ij}} = \sum_n \frac{\partial \mathcal{L}}{\partial i_{in_j}^l[n]} \cdot v_{sp_i}^{l-1}[n]
\end{align}
\subsubsection{$v_{m}[n]$}
The forward pass equation for computing $v_{m}[n]$ is given by (\ref{eq:difference_equation_lif}). All the corresponding $v_m$, $v_{sp}$, and $i_{in}$ are $v_{m_i}^l$, $v_{sp_i}^l$, and $i_{in_i}^l$, respectively. Here, $v_{sp_i}^l$ acts as the gating function to determine whether the input current and membrane potential of the previous timestep should be used to compute the membrane potential of the current timestep. The backward pass equations for computing the gradients with respect to $i_{in_i}^l[n]$, $v_{m_i}^l[n-1]$, and $i_{in_i}^l[n-1]$ are given by (\ref{eq:backward_pass_local_vm_iinn}), (\ref{eq:backward_pass_local_vm_vm}), and (\ref{eq:backward_pass_local_vm_iin_1}), respectively.
\begin{align}
    \label{eq:backward_pass_local_vm_iinn}
    \frac{\partial v_{m_i}^l[n]}{\partial i_{in_i}^l[n]} &= \frac{1}{c + \lambda} \\
    \label{eq:backward_pass_local_vm_vm}
    \frac{\partial v_{m_i}^l[n]}{\partial v_{m_i}^l[n-1]} &= \bar{v}_{sp_i}^l[n-1] \cdot \frac{c - \lambda}{c + \lambda} \\
    \label{eq:backward_pass_local_vm_iin_1}
    \frac{\partial v_{m_i}^l[n]}{\partial i_{in_i}^l[n-1]} &= \bar{v}_{sp_i}^l[n-1] \cdot \frac{1}{c + \lambda}
\end{align}
The gradient with respect to the loss function is given by (\ref{eq:backward_pass_global_iin}) using the chain rule of differentiation.
\begin{equation}
    \label{eq:backward_pass_global_iin}
    \frac{\partial \mathcal{L}}{\partial i_{in_i}^l[n]} = \frac{1}{c + \lambda} \cdot \left( \frac{\partial \mathcal{L}}{\partial v_{m_i}^l[n]} + \bar{v}_{sp_i}^l[n] \cdot \frac{\partial \mathcal{L}}{\partial v_{m_i}^l[n+1]} \right)
\end{equation}
For the output layer, the forward pass is given by (\ref{eq:forward_pass_vm_output}).
\begin{align}
    \label{eq:forward_pass_vm_output}
    v_{m_i}^L[n] &= i_{in_i}^L[n] + i_{in_i}^L[n-1] + v_{m_i}^L[n-1] \nonumber \\
    v_{m_i}^L[T-1] &= \left( 2 \cdot \sum_n i_{in_i}^L[n] \right) - i_{in_i}^L[T-1]
\end{align}
The gradient is given by (\ref{eq:backward_pass_global_iin_output}).
\begin{align}
    \label{eq:backward_pass_global_iin_output}
    \frac{\partial \mathcal{L}}{\partial i_{in_i}^L[n]} &= k \cdot \frac{\partial \mathcal{L}}{\partial v_{m_i}^L[T-1]} \\
    k &= \begin{cases}
        1, & \text{if } n = T-1 \\
        2, & \text{otherwise}
    \end{cases} \nonumber
\end{align}
The relation between $v_{m_i}^L[T-1]$ and $\mathcal{L}$ is given by (\ref{eq:loss_function}) and the derivative is given by (\ref{eq:loss_function_derivative}).
\begin{equation}
    \label{eq:loss_function_derivative}
    \frac{\partial \mathcal{L}}{\partial v_{m_i}^L[T-1]} = \frac{e^{v_{m_i}^L[T-1]}}{\sum_{j}e^{v_{m_j}^L[T-1]}} - y_i
\end{equation}
\subsubsection{$v_{sp}[n]$}
\begin{figure}[htbp]
    \centering
    \vspace{-1em}
    \includegraphics[scale=0.78]{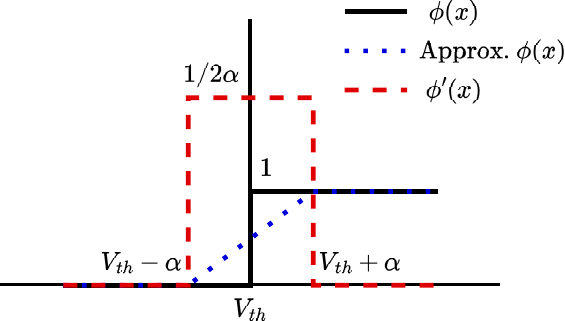} \vspace{-0.5em}
    \caption{Approximation of the thresholding function as linear in the region around the threshold value.} \vspace{-1em}
    \label{fig:thresholding_function}
\end{figure}

When the membrane potential crosses a threshold value, a spike is produced. The forward pass equation for computing $v_{sp}[n]$ is given by (\ref{eq:forward_pass_vsp}). \vspace{-1em}
\begin{equation}
    \label{eq:forward_pass_vsp}
    v_{sp_i}^l[n] = \phi(v_{m_i}^l[n]) = \begin{cases}
        1, & \text{if } v_{m_i}^l[n] \geq V_{th} \\
        0, & \text{otherwise}
    \end{cases} \vspace{-0.5em}
\end{equation}
The thresholding activation function, $\phi(x)$, that produces the spike is non-differentiable. To ensure that the gradient is propagated through the thresholding function, the function is approximated to be a linear function in the region around the threshold value, as shown in Fig. \ref{fig:thresholding_function} and described in \cite{Yujie2018}. The gradient of the approximated function is given by (\ref{eq:thresholding_function_derivative}). The variable $\alpha$ in (\ref{eq:thresholding_function_derivative}) and Fig. \ref{fig:thresholding_function} is a hyperparameter that can be tuned. For simplicity, $\alpha$ is set to 0.5 in this work. \vspace{-0.5em}
\begin{equation}
    \label{eq:thresholding_function_derivative}
    \frac{\partial v_{sp_i}^l[n]}{\partial v_{m_i}^l[n]} = \phi'(v_{m_i}^l[n]) = \begin{cases}
        \frac{1}{2 \alpha}, & \text{if } |v_{m_i}^l[n] - V_{th}| \leq \alpha \\
        0, & \text{otherwise}
    \end{cases}
\end{equation}

\begin{figure}[htbp]
    \centering
    \vspace{-1em}
    \includegraphics[scale=0.18]{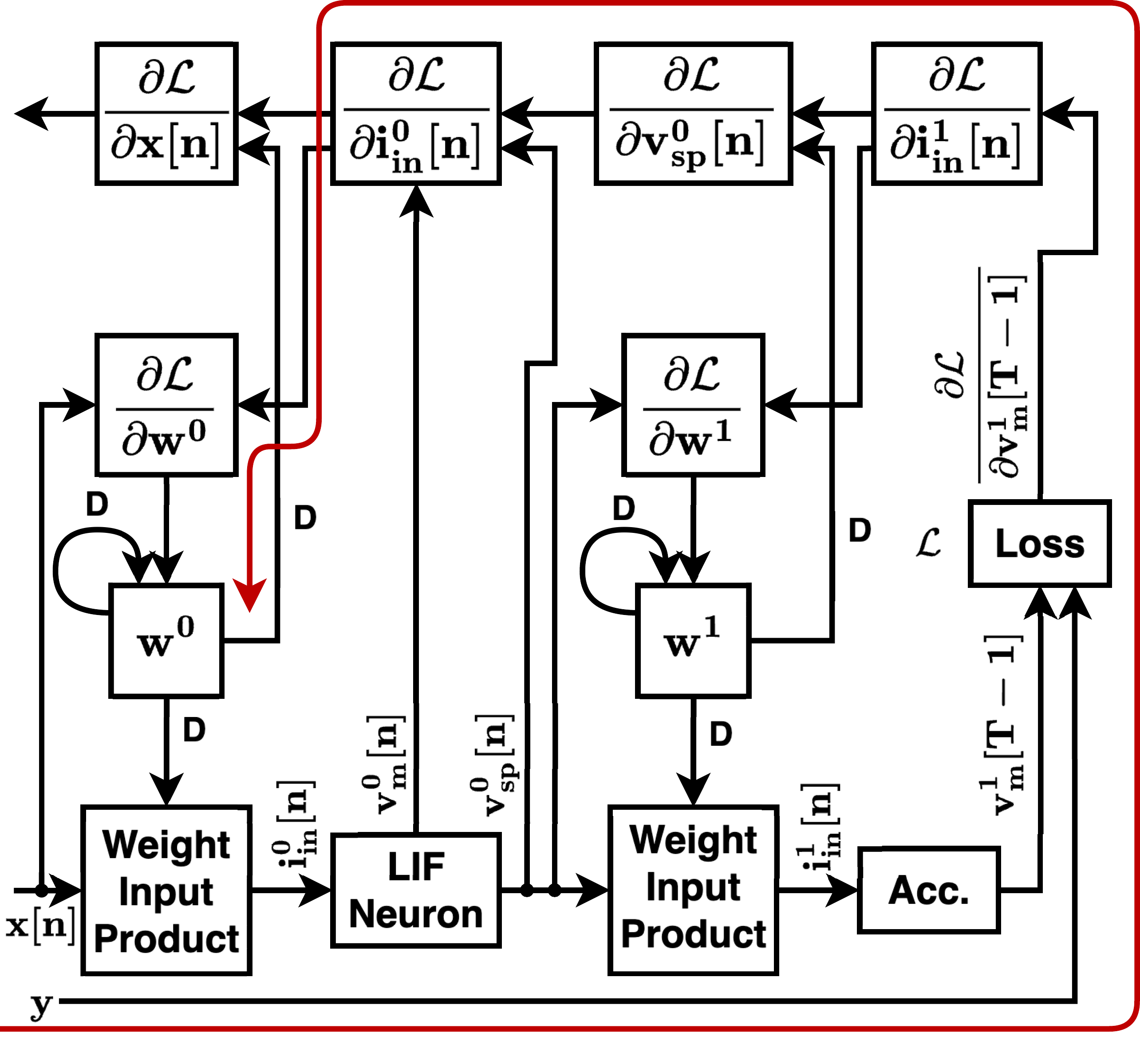} \vspace{-1em}
    \caption{Dataflow graph of training a sample two-layer network. The colored line indicates the critical loop.} \vspace{-1em}
    \label{fig:training_dfg}
\end{figure}

\begin{figure}[htbp]
    \centering
    \vspace{-1em}
    \includegraphics[scale=0.8]{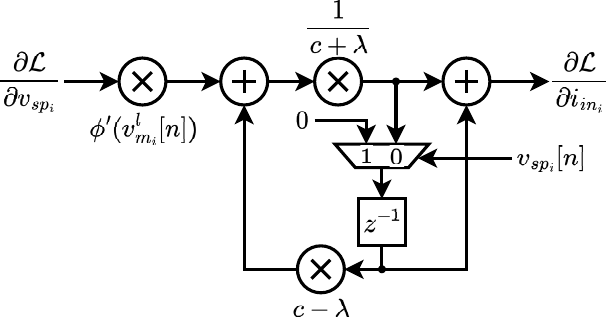} \vspace{-1em}
    \caption{Structure of the IIR filter for computing the gradient of the loss with respect to the input current.} \vspace{-1em}
    \label{fig:gradient_iir_filter}
\end{figure}

For all layers except the output layer, the gradient of the loss with respect to membrane potential is given by (\ref{eq:backward_pass_global_vm}).
\begin{align}
    \label{eq:backward_pass_global_vm}
    \frac{\partial \mathcal{L}}{\partial v_{m_i}^l[n]} &= \frac{\partial \mathcal{L}}{\partial v_{sp_i}^l[n]} \cdot \frac{\partial v_{sp_i}^l[n]}{\partial v_{m_i}^l[n]} + \frac{\partial \mathcal{L}}{\partial v_{m_i}^l[n+1]} \cdot \frac{\partial v_{m_i}^l[n+1]}{\partial v_{m_i}^l[n]} \nonumber \\
    &= \frac{\partial \mathcal{L}}{\partial v_{sp_i}^l[n]} \cdot \phi'(v_{m_i}^l[n]) \nonumber \\ &\mathrel{\phantom{=}}+ \frac{\partial \mathcal{L}}{\partial v_{m_i}^l[n+1]} \cdot \bar{v}_{sp_i}^l[n] \cdot \frac{c - \lambda}{c + \lambda}
\end{align}
From (\ref{eq:backward_pass_global_iin}) and (\ref{eq:backward_pass_global_vm}), it is apparent that the gradient of the loss with respect to the input current can be computed using a similar IIR filter structure as the LIF neuron model. The structure of the IIR filter for the gradient computation is shown in Fig. \ref{fig:gradient_iir_filter}. All the variables in the structure are in time-reversed order, \emph{i.e.,} $\frac{\partial \mathcal{L}}{\partial v_{sp_i}[T-1]}$ is the input in the first clock cycle, $\frac{\partial \mathcal{L}}{\partial v_{sp_i}[T-2]}$ in the second clock cycle, and so on. Similarly, the output is also produced in time-reversed order.

 \vspace{-1em}
\subsection{Dataflow Graph of Training}

Figure \ref{fig:training_dfg} shows the DFG for training a sample two-layer network. Each node in the DFG computes the corresponding variables for a single mini-batch. Therefore, one weight update based on a single input mini-batch is computed per cycle of the DFG. The DFG is similar to the DFG of a conventional neural network, the difference being each computation in the DFG of an SNN has a timestep dimension. The training process inherently has feedback loops. The colored line in Fig. \ref{fig:training_dfg} highlights the critical loop in the example DFG, which has the forward and backward computations of all the layers. Since there is only one delay element in the critical loop, it is not straightforward to retime the DFG for mapping onto multiple processors. \vspace{-1em}
\subsection{Delayed Gradients}
\label{sec:delayed_gradients}
\begin{figure}[htbp]
    \centering
    \includegraphics[scale=0.78]{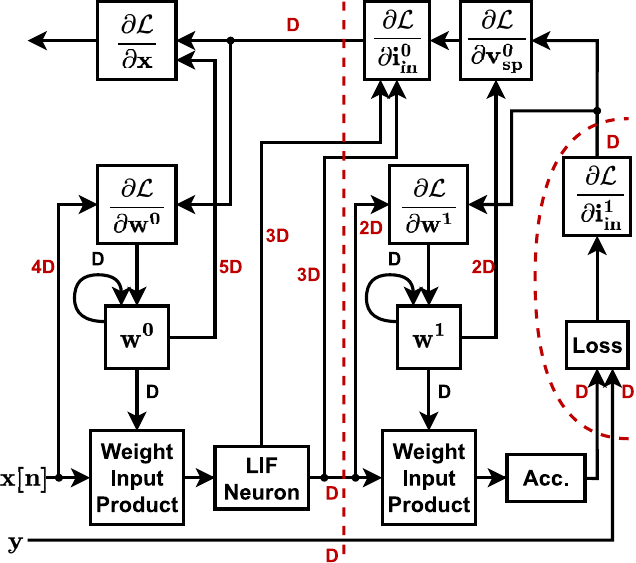} \vspace{-1em}
    \caption{Retimed DFG with delayed gradients. The DFG can be split into subgraphs at the colored lines. The colored delay elements emerge from retiming the DFG using delayed gradients.} \vspace{-1em}
    \label{fig:retimed_dfg}
\end{figure}

To split the DFG into multiple subgraphs and map onto multiple processors, additional delay elements must be added into the critical loop. One way of achieving this is using delayed gradients \cite{Long1989}. The gradient update is delayed by a few cycles, introducing additional delay elements into the critical loop. The modified DFG can now be retimed and split into multiple subgraphs. The use of delayed gradients and retiming was used to achieve inter-layer pipelining in the training of the CNNs in the LayerPipe approach described in \cite{Unnikrishnan2021}. Figure \ref{fig:retimed_dfg} shows the DFG after retiming with delayed gradients.

The introduction of delayed gradients calls for an analysis of the trainability and convergence of the network. Delayed gradients will not severely affect the trainability of the network if the gradients change slowly over iterations. In the context of neural networks, this is determined by the gradient update algorithm used to train the network. The most common gradient update algorithm is the stochastic gradient descent (SGD) algorithm. The SGD algorithm is given by (\ref{eq:sgd_update}), where $w$ is the weight, $\eta$ is the learning rate, and $\frac{\partial \mathcal{L}}{\partial w}$ is the gradient of the loss with respect to the weight.
\begin{equation}
    \label{eq:sgd_update}
    w_{t+1} = w_t - \eta \cdot \frac{\partial \mathcal{L}}{\partial w}
\end{equation}
From the SGD update equation, the gradients may change rapidly in every iteration during the initial stages of training. This would result in delayed gradients significantly affecting the trainability of the network. The effect of delayed gradients on the convergence of LMS filters is explored in \cite{Shanbhag1993} by using the moving average of the gradients. In the context of neural networks, the Adam optimizer \cite{Kingma2014} ensures a similar behavior by introducing first and second-order momentum. The Adam optimizer is given by (\ref{eq:adam_update}).
\begin{align}
    \label{eq:adam_update}
    m_{t+1} &= \beta_1 \cdot m_t + (1 - \beta_1) \cdot \frac{\partial \mathcal{L}}{\partial w} \nonumber \\
    v_{t+1} &= \beta_2 \cdot v_t + (1 - \beta_2) \cdot \left( \frac{\partial \mathcal{L}}{\partial w} \right)^2 \nonumber \\
    \eta_{t+1} &= \eta \cdot \frac{\sqrt{1 - \beta_2^{t+1}}}{1 - \beta_1^{t+1}} \nonumber \\
    w_{t+1} &= w_t - \eta_{t+1} \cdot \frac{m_{t+1}}{\sqrt{v_{t+1}} + \epsilon}
\end{align}
\subsection{Example Networks}
To analyze the effect of delayed gradients, the MNIST \cite{LeCun1998}, Neuromorphic-MNIST \cite{Orchard2015}, and DVS128 Gestures \cite{Amir2017} datasets are considered. Table \ref{tab:example_network} shows the networks used for the three datasets which are trained for 8, 30, and 40 timesteps, respectively. The networks are trained using the SGD and Adam optimizers, with and without delayed gradients. The gradients in the layers are delayed by a predetermined number of batches, as tabulated in Table \ref{tab:example_network}. This distribution of delays splits the training DFG into subgraphs such that there is only one layer per subgraph, as described in \cite{Unnikrishnan2021}. The learning rate is set to $0.001$ for both optimizers. The mini-batch size is set to $32$ and the MNIST and N-MNIST networks are trained for $100$ epochs, while the DVS128 Gestures network is trained for $300$ epochs. The results are shown in Table \ref{tab:sgd_adam_delayed_results}. The mean and standard deviation of the accuracy for $10$ different weight initializations are reported.
\begin{table}[htbp]
    \renewcommand{\arraystretch}{1.1}
    \centering
    \caption{Details of the networks used for the three datasets. The delay column indicates the number of batches by which the gradients of the corresponding layers are delayed.} \vspace{-1em}
    \label{tab:example_network}
    \begin{tabular}{|c|c|c|c|c|}
        \hline
        \hline
        \textbf{Dataset} & \textbf{Layer} & \textbf{Output Shape} & \textbf{\# Params} & \textbf{Delay}\\
        \hline
        \hline
        \multirow{6}{*}{MNIST} & Conv 1 & $(28 \times 28 \times 8)$ & 80 & 6 \\
        \cline{2-5}
        & Maxpool & $(14 \times 14 \times 8)$ & - & - \\
        \cline{2-5}
        & Conv 2 & $(14 \times 14 \times 8)$ & 584 & 4 \\
        \cline{2-5}
        & Maxpool & $(7 \times 7 \times 8)$ & - & - \\
        \cline{2-5}
        & FC 1 & $128$ & 50,304 & 2 \\
        \cline{2-5}
        & Output & $10$ & 1,290 & 0 \\
        \hline
        \hline

        \multirow{6}{*}{N-MNIST} & Conv 1 & $(32 \times 32 \times 8)$ & 152 & 6 \\
        \cline{2-5}
        & Maxpool & $(16 \times 16 \times 8)$ & - & - \\
        \cline{2-5}
        & Conv 2 & $(16 \times 16 \times 8)$ & 584 & 4 \\
        \cline{2-5}
        & Maxpool & $(8 \times 8 \times 8)$ & - & - \\
        \cline{2-5}
        & FC 1 & $32$ & 16,416 & 2 \\
        \cline{2-5}
        & Output & $10$ & 330 & 0 \\
        \hline
        \hline

        \multirow{12}{*}{\Centerstack{DVS128 Gestures}} & Conv 1 & $(64 \times 64 \times 32)$ & 608 & 14 \\
        \cline{2-5}
        & Maxpool & $(32 \times 32 \times 32)$ & - & - \\
        \cline{2-5}
        & Conv 2 & $(32 \times 32 \times 64)$ & 18,496 & 14 \\
        \cline{2-5}
        & Maxpool & $(16 \times 16 \times 64)$ & - & - \\
        \cline{2-5}
        & Conv 3 & $(16 \times 16 \times 128)$ & 73,856 & 12 \\
        \cline{2-5}
        & Conv 4 & $(16 \times 16 \times 128)$ & 147,584 & 8 \\
        \cline{2-5}
        & Maxpool & $(8 \times 8 \times 128)$ & - & - \\
        \cline{2-5}
        & Conv 5 & $(8 \times 8 \times 256)$ & 295,168 & 4 \\
        \cline{2-5}
        & Conv 6 & $(8 \times 8 \times 256)$ & 590,080 & 2 \\
        \cline{2-5}
        & Maxpool & $(4 \times 4 \times 256)$ & - & - \\
        \cline{2-5}
        & FC 1 & $128$ & 524,416 & 0 \\
        \cline{2-5}
        & Output & $11$ & 2,827 & 0 \\
        \hline
        \hline
    \end{tabular}
    \vspace{-1em}
\end{table}

\begin{table}[htbp]
    \renewcommand{\arraystretch}{1.2}
    \centering
    \caption{Accuracy of the networks when trained with SGD and Adam optimizers, with and without delayed gradients. Mean and standard deviation of accuracies over 10 different weight initializations are reported.} \vspace{-1em}
    \label{tab:sgd_adam_delayed_results}
    \begin{tabular}{|c|c|c|c|c|}
        \hline
        \hline
        \multirow{2}{*}{\textbf{Dataset}} & \multirow{2}{*}{\textbf{Optimizer}} & \multicolumn{2}{c|}{\textbf{Delayed Gradients}} \\
        \cline{3-4}
        & & \textbf{No} & \textbf{Yes} \\
        \hline
        \hline
        \multirow{2}{*}{MNIST} & SGD & $96.37\% \pm 0.32\%$ & $96.38\% \pm 0.25\%$ \\
        \cline{2-4}
        & Adam & $98.64\% \pm 0.13\%$ & $98.59\% \pm 0.05\%$ \\
        \hline
        \hline
        \multirow{2}{*}{N-MNIST} & SGD & $96.54\% \pm 0.32\%$ & $96.34\% \pm 0.52\%$ \\
        \cline{2-4}
        & Adam & $98.17\% \pm 0.15\%$ & $98.13\% \pm 0.12\%$ \\
        \hline
        \hline
        \multirow{2}{*}{\Centerstack{DVS128 Gestures}} & SGD & $79.03\% \pm 5.71\%$ & $73.35\% \pm 2.71\%$ \\
        \cline{2-4}
        & Adam & $85.42\% \pm 2.47\%$ & $76.39\% \pm 3.52\%$ \\
        \hline
        \hline
    \end{tabular}
    \vspace{-2em}
\end{table}

It is observed that the smaller networks trained on the MNIST and N-MNIST datasets achieve similar accuracy with delayed gradients as without delayed gradients for both SGD and Adam optimizers. As expected, the accuracy with the Adam optimizer is higher than that of the SGD optimizer.

However, for the larger network trained on the DVS128 Gestures dataset, delayed gradients impact the performance. It is also observed that the difference in accuracy in the networks trained with and without delayed gradients increases as the absolute accuracy obtained increases. Therefore, delayed gradients have a more significant effect on the trainability of the network when the network is trained to achieve higher accuracy, as is the case with the Adam optimizer, as opposed to the SGD optimizer. However, the maximum accuracy achieved across the 10 runs in Adam is $81.25\%$, which is higher than the $77.43\%$ achieved with SGD. Therefore, the Adam optimizer is used for training the networks in the rest of the work. It is to be noted that the networks trained with delayed gradients offer a significant speedup in training time, as discussed in Section \ref{sec:results}, at the cost of a small reduction in accuracy. The trained networks can be further fine-tuned without delayed gradients to achieve higher accuracy if required. However, such fine-tuning is not explored in this work.

The obtained accuracies are compared against prior SNN works which have targeted the same three datasets, and tabulated in Table \ref{tab:accuracy_comparison}. The networks trained using the Adam optimizer with delayed gradients have comparable accuracy to the prior works on the MNIST and N-MNIST datasets. The accuracy on the DVS128 Gestures dataset is lower than the prior works. However, the network trained in this work has fewer parameters and timesteps than the prior works. Moreover, the training method used in this work is a generic method without any dataset-specific tuning such as data preprocessing. There is also no fine-tuning done without delayed gradients to further improve the accuracy. Therefore, the obtained accuracy is reasonable.
\begin{table}
    \renewcommand{\arraystretch}{1.2}
    \centering
    \caption{Comparison of the accuracies obtained in this work with prior works.} \vspace{-1em}
    \label{tab:accuracy_comparison}
    \begin{tabular}{|c|c|c|c|c|}
        \hline
        \hline
        \textbf{Work} & \textbf{Dataset} & \textbf{Accuracy (\%)} & \textbf{\# Params} & \textbf{Timesteps} \\
        \hline
        \hline
        \cite{Yujie2018} & \multirow{6}{*}{MNIST} & 98.89 & 636,010 & 30 \\
        \cline{1-1} \cline{3-5}
        \cite{Yujie2018} & & 99.42 & 606,740 & 30 \\
        \cline{1-1} \cline{3-5}
        \cite{Pfeiffer2016} & & 99.31 & 517,780 & - \\
        \cline{1-1} \cline{3-5}
        \cite{Stromatias2017} & & 98.42 & 22,662 & - \\
        \cline{1-1} \cline{3-5}
        \cite{Lee2020} & & 99.59 & 517,780 & 50 \\
        \cline{1-1} \cline{3-5}
        \textbf{This Work} & & \textbf{98.46} & \textbf{52,258} & \textbf{8} \\
        \hline
        \hline
        \cite{Yujie2018} & \multirow{5}{*}{N-MNIST} & 98.78 & 1,858,410 & 30 \\
        \cline{1-1} \cline{3-5}
        \cite{Pfeiffer2016} & & 98.66 & 1,858,410 & - \\
        \cline{1-1} \cline{3-5}
        \cite{Stromatias2017} & & 97.23 & 37,044 & - \\
        \cline{1-1} \cline{3-5}
        \cite{Lee2020} & & 99.09 & 750,780 & 50 \\
        \cline{1-1} \cline{3-5}
        \textbf{This Work} & & \textbf{98.06} & \textbf{17,482} & \textbf{30} \\
        \hline
        \hline
        \cite{Zhu2022} & \multirow{3}{*}{\Centerstack{DVS128 Gestures}} & 95.83 & 2,081,866 & 500 \\
        \cline{1-1} \cline{3-5}
        \cite{He2020} & & 93.40 & 2,332,891 & - \\
        \cline{1-1} \cline{3-5}
        \textbf{This Work} & & \textbf{81.25} & \textbf{1,653,035} & \textbf{40} \\
        \hline
        \hline
    \end{tabular}
    \vspace{-1em}
\end{table} \vspace{-1em}
\section{Accelerators for Training SNNs}
\begin{figure}[htbp]
    \centering
    \vspace{-1em}
    \includegraphics[scale=0.8]{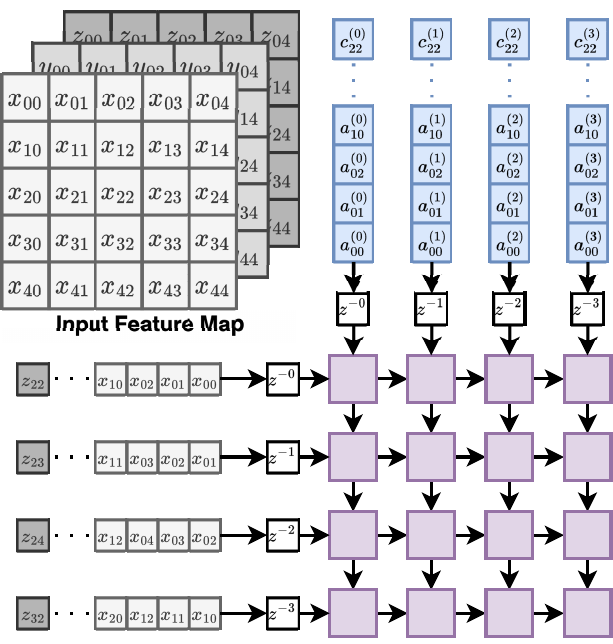} \vspace{-1em}
    \caption{Structure of a regular $4 \times 4$ systolic array in output-stationary configuration for computing a convolution. The input is a $5 \times 5$ feature map with $3$ channels, which is convolved with four $3 \times 3$ filters. $a^{(f)}_{ij}$ is the coefficient of filter $f$ at the $i,j^{th}$ location for the first channel. The second and third channels are represented by $b^{(f)}_{ij}$ and $c^{(f)}_{ij}$, respectively.}
    \label{fig:regular_systolic_array} \vspace{-1em}
\end{figure}

Recent years have seen significant progress in developing efficient hardware accelerators for training neural networks. Most accelerators developed for training neural networks are based on systolic arrays \cite{Kung1978, Parhi1999, Wei2017}, like the Google Tensor Processing Unit (TPU) \cite{Jouppi2017}. Systolic arrays are well-suited for training neural networks because of their regular structure and high compute density. Figure \ref{fig:regular_systolic_array} shows the structure of a typical $4 \times 4$ systolic array. Each element in the array is a processing element (PE) that performs a single multiply-and-accumulate (MAC) operation every cycle. The structure in Fig. \ref{fig:regular_systolic_array} is an output-stationary architecture performing the convolution operation. The input is a $5 \times 5$ feature map with $3$ channels, which is convolved with four $3 \times 3$ filters. The output is a $3 \times 3$ feature map with $4$ channels. The first element in every channel of the output feature map is computed by multiplying the first $3 \times 3$ patch of the input feature map with the four filters. This operation is done in the first row of the systolic array, where the $3 \times 3$ patch of the input feature map is streamed through the row and the filters are streamed down the four columns. In the structure shown in Fig. \ref{fig:regular_systolic_array}, only four elements of the output feature map can be computed per channel as there are only four rows in the systolic array. The remaining elements of the output feature map are computed after the first four are computed and streamed out.
\vspace{-1em}
\subsection{Modified systolic array for SNNs}
\begin{figure}[htbp]
    \centering
    \vspace{-1em}
    \includegraphics[scale=0.8]{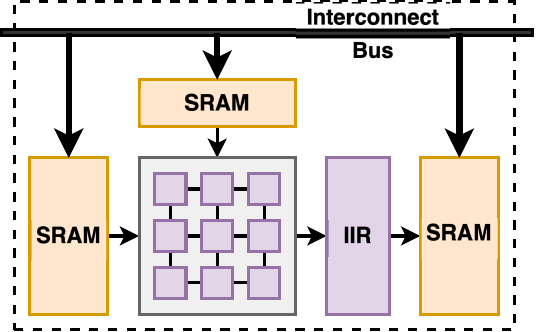} \vspace{-1em}
    \caption{Modified systolic array processor for training SNNs. The interconnect bus connects adjacent processors and facilitates data transfer between them.} \vspace{-1em}
    \label{fig:iir_systolic_array}
\end{figure}

The systolic array structure and the data streaming scheme need to be modified to train SNNs. Both the forward pass and backward pass of the LIF neuron model require an IIR filter. Therefore, a bank of IIR filters should be present along with the systolic array. The output of each column of the systolic array is passed through an IIR filter and sent to the output SRAM. The number of IIR filters in the filter bank is $S_C$, where $S_C$ is the number of columns in the systolic array. Figure \ref{fig:iir_systolic_array} shows the architecture of the modified systolic array processor for SNNs. The interconnect bus in Fig. \ref{fig:iir_systolic_array} connects adjacent processors and facilitates data transfer between them.

\subsection{Estimating clock cycles for computation}

The discussed processor has to be modeled to accurately estimate the number of clock cycles required to compute a given task \cite{Samajdar2020}. Since the target problem is image classification, only two types of computations are required: convolution and matrix multiplication. For each of these, there are three cases to consider: the forward pass, gradient with respect to weights, and gradient with respect to inputs. The mapping of the computations to the systolic array for each of these cases is discussed in this section.

\subsubsection{Convolutional layer}

In the forward pass of convolution, the input image is convolved with a set of filters. Assuming the image to be $H \times W \times C$ including padding, and the filter to be $K \times K \times C$, the output of the convolution is $H_{out} \times W_{out} \times 1$, where $H_{out} = (H - K + 1)$ and $W_{out} = (W - K + 1)$. The convolution operation is performed by sliding the filter over the input image and computing the dot product of the filter and the image patch. The dot product is computed by multiplying the corresponding elements of the filter and the image patch and summing them up. This is repeated for $T$ timesteps and $F$ filters.

For each of the three tasks, \emph{i.e.,} the forward pass, gradient with respect to weights, and gradient with respect to inputs, the number of rows and columns necessary for computation, and the number of MACs per PE are different. Table \ref{tab:conv_tasks} summarizes the number of rows and columns and the number of MACs per PE for each of the three tasks. In most practical cases, the systolic arrays are not large enough to accommodate all the rows and columns necessary for computation. Therefore, the computations are tiled to fit the systolic array. The number of tiles required in an $(S_R \times S_C)$ systolic array is given by (\ref{eq:n_tiles}).
\begin{equation}
    \label{eq:n_tiles}
    N_{tiles} = \left\lceil \frac{N_{rows}}{S_R} \right\rceil \cdot \left\lceil \frac{N_{cols}}{S_C} \right\rceil
\end{equation}
\noindent where, $\lceil x \rceil$ is the ceiling function. For each tile, the number of clock cycles is the number of cycles required for accumulating the output, and the input and output skews, given by (\ref{eq:conv_cycles_tile}). \vspace{-0.5em}
\begin{equation}
    \label{eq:conv_cycles_tile}
    N_{cycles_{tile}} = N_{MAC} + (S_R - 1) + (S_C - 1)
\end{equation}
The total number of clock cycles required for each task is given by (\ref{eq:conv_cycles_forward}) when $N_{tiles}$ and $N_{cycles_{tile}}$ are computed using the values from Table \ref{tab:conv_tasks}. \vspace{-0.5em}
\begin{equation}
    \label{eq:conv_cycles_forward}
    N_{conv} = N_{tiles} \cdot N_{cycles_{tile}}
\end{equation} \vspace{-1em}
\begin{table}[htbp]
    \renewcommand{\arraystretch}{1.1}
    \centering
    \vspace{-1em}
    \caption{Number of rows and columns, and number of MACs per PE for each of the three tasks for convolution.} \vspace{-1em}
    \label{tab:conv_tasks}
    \begin{tabular}{|c|c|c|c|}
        \hline
        \hline
        \textbf{Task} & $N_{rows}$ & $N_{cols}$ & $N_{MAC}$ \\
        \hline
        \hline
        Forward pass & $H_{out} \cdot W_{out} \cdot T$ & $F$ & $K \cdot K \cdot C$ \\
        \hline
        Weight Gradient & $K \cdot K \cdot C$ & $F$ & $H_{out} \cdot W_{out} \cdot T$ \\
        \hline
        Input Gradient & $H \cdot W \cdot T$ & $C$ & $K \cdot K \cdot F$ \\
        \hline
        \hline
    \end{tabular} \vspace{-1em}
\end{table}
\subsubsection{Fully-connected layer}
In the forward pass of a fully-connected layer simulated over $T$ timesteps, the input feature map is a vector of size $Q_{in}$, and the output feature map is a vector of size $Q_{out}$. The number of tiles and number of clock cycles per tile are given by (\ref{eq:n_tiles}) and (\ref{eq:conv_cycles_tile}), respectively. Similar to the convolutional layer, the total number of clock cycles required for the fully-connected layer is given by (\ref{eq:fc_cycles}) when $N_{tiles}$ and $N_{cycles_{tile}}$ are computed using the values from Table \ref{tab:fc_tasks}.
\begin{equation}
    \label{eq:fc_cycles}
    N_{fc} = N_{tiles} \cdot N_{cycles_{tile}}
\end{equation} \vspace{-1em}
\begin{table}[htbp]
    \renewcommand{\arraystretch}{1.1}
    \centering
    \vspace{-1em}
    \caption{Values of $N_R$, $N_C$, and $N_{MAC}$ for each of the tasks in the fully-connected layer.} \vspace{-1em}
    \label{tab:fc_tasks}
    \begin{tabular}{|c|c|c|c|}
        \hline
        \hline
        \textbf{Task} & $N_{rows}$ & $N_{cols}$ & $N_{MAC}$ \\
        \hline
        \hline
        Forward pass & $T$ & $Q_{out}$ & $Q_{in}$ \\
        \hline
        Weight Gradient & $Q_{in}$ & $Q_{out}$ & $T$ \\
        \hline
        Input Gradient & $T$ & $Q_{in}$ & $Q_{out}$ \\
        \hline
        \hline
    \end{tabular} \vspace{-1em}
\end{table} \vspace{-1em}
\subsection{Clock cycles for the MNIST network}
For the four-layer network trained on MNIST detailed in Table \ref{tab:example_network}, the number of clock cycles required for each layer when mapped to a $32 \times 32$ systolic array is given in Table \ref{tab:example_network_cycles}. For most practical applications, the input gradient of the first layer is not required. Therefore, the total number of clock cycles needed for one weight update of the entire network is $46,956$. \vspace{-1em}

\begin{table}[htbp]
    \renewcommand{\arraystretch}{1.1}
    \centering
    \vspace{-1em}
    \caption{Number of clock cycles required for each layer of the MNIST network when mapped to a $32 \times 32$ systolic array.} \vspace{-1em}
    \label{tab:example_network_cycles}
    \begin{tabular}{|c|c|c|c|}
        \hline
        \hline
        \textbf{Layer} & \textbf{Forward pass} & \textbf{Weight gradient} & \textbf{Input gradient} \\
        \hline
        \hline
        Conv1 & 13,916 & 6,334 & 26,264 \\
        \hline
        Conv2 & 6,566 & 4,890 & 6,566 \\
        \hline
        FC1 & 1,816 & 3,640 & 2,470 \\
        \hline
        Output & 190 & 280 & 288 \\
        \hline
        \hline
    \end{tabular} \vspace{-1em}
\end{table} \vspace{-1em}
\section{Scheduling to Multiple Processors}
When the training is done with more than one processor, the workload should be distributed evenly among the processors to minimize the total number of clock cycles required for the entire network. Ideally, the training throughput should be increased by a factor equal to the number of processors used. However, achieving this in practice is difficult. This section explores various scheduling algorithms that can distribute the workload among multiple processors and compares their performance in terms of throughput. The number of cycles required for the forward pass, weight gradient, and input gradient of a layer indexed $l$ is denoted by $N_{FP}^l$, $N_{WG}^l$, and $N_{IG}^l$, respectively. The number of processors used for training is denoted by $P$. The total number of clock cycles required for the entire network is represented by $N_{total}$. The network trained on MNIST detailed in Table \ref{tab:example_network} is used for the simulations in this section. For this network, $N_{total}$ is 46,956 clock cycles. The values of $N_{FP}^l$, $N_{WG}^l$, and $N_{IG}^l$, for all $l$, are tabulated in Table \ref{tab:example_network_cycles}. $N_{IG}^1$ corresponds to the computation of $\mathbf{\partial \mathcal{L}/\partial x[n]}$ in Fig. \ref{fig:training_dfg}, and is ignored in the simulations since it is not required for training.

The basis for all the discussed scheduling algorithms is the ability to split the training dataflow graph shown in Fig. \ref{fig:training_dfg} into various subgraphs. As discussed in Section \ref{sec:delayed_gradients}, this can be achieved with the help of delayed gradients. The DFG can be retimed and split into subgraphs, shown by the dashed lines in Fig. \ref{fig:retimed_dfg}. Each subgraph can be mapped onto a processor. The number of batches by which each batch is delayed depends on the processor to which the corresponding weight gradient computation task is mapped. For each weight gradient task, the number of delays would be twice the number of processors following the processor to which the task is mapped. For example, if a weight gradient task is mapped to the third processor, and there are a total of $5$ processors, the number of delays would be $2 \times (5 - 3) = 4$.

Throughout this section, the speedup of training is compared to the single processor implementation. If the training is done on a single processor, the number of clock cycles per weight update is $N_{total}$, or $46,956$ for the network under consideration. The number of required clock cycles reduces when the training is done on multiple processors. The speedup of training is the ratio of the number of clock cycles per weight update on a single processor to the number of clock cycles per weight update on multiple processors. \vspace{-1em}
\subsection{Layer-wise scheduling}
\begin{figure}[htbp]
    \centering
    \vspace{-1em}
    \includegraphics[scale=0.8]{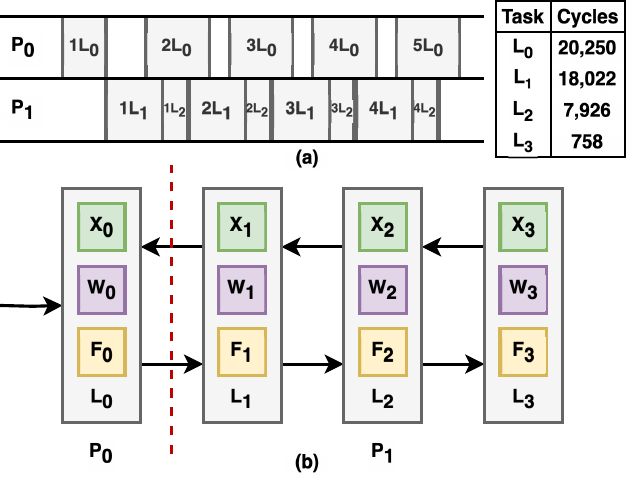} \vspace{-1em}
    \caption{(a) Schedule map for the layer-wise scheduling algorithm when $P = 2$ and (b) split of the DFG to two subgraphs. $xL_i$ represents the tasks of layer $i$ for input batch index $x$.} \vspace{-1em}
    \label{fig:layerwise_map}
\end{figure}

The most straightforward way to distribute the workload among multiple processors is to consider an entire layer as a single task and schedule the tasks to the processors. With this approach, the speedup in computation, $\sigma$, is upper bounded by the number of clock cycles required for the longest layer, as given by (\ref{eq:layerwise_speedup_bound}).
\begin{equation}
    \label{eq:layerwise_speedup_bound}
    \sigma_{layerwise} \leq \frac{N_{total}}{\max_l(N_{FP}^l + N_{WG}^l + N_{IG}^l)}
\end{equation}
For the network trained on MNIST detailed in Table \ref{tab:example_network}, the longest layer is the first layer, with $20,250$ clock cycles\footnote{Note that the input gradient task is ignored for the first layer.}. $\sigma_{layerwise}$ for this network is $2.32$. Therefore, using more than three processors does not improve training throughput. Figure \ref{fig:layerwise_map}(a) shows the schedule map for the layer-wise scheduling algorithm when $P = 2$ along with the number of clock cycles necessary for each task. The schedule map shows the processor to which each task is assigned. After the pipeline is full, the processor $P_0$ is active for $20,250$ cycles and the processor $P_1$ is active for $26,706$ cycles. The number of cycles per weight update is $26,706$, resulting in a speedup of $1.8\times$ over the single processor implementation. Figure \ref{fig:layerwise_map}(b) shows the split of the DFG into two subgraphs that are mapped to the corresponding processors.

 \vspace{-1em}
\subsection{PipeDream-based scheduling}
\begin{figure}[htbp]
    \centering
    \vspace{-1em}
    \includegraphics[scale=0.8]{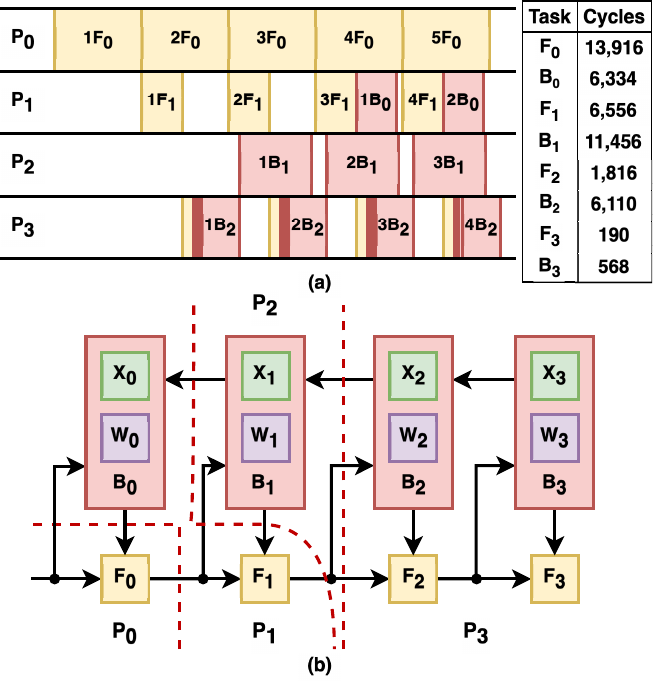} \vspace{-1em}
    \caption{(a) Schedule map for the PipeDream-based scheduling algorithm when $P = 4$ and (b) split of the DFG to four subgraphs.} \vspace{-1em}
    \label{fig:pipedream_map}
\end{figure}

Scheduling based on the PipeDream algorithm \cite{Narayanan2019} assumes the layer's tasks are split into forward and backward passes. The backward pass contains both input gradient computation and weight gradient computation. This method of splitting increases the upper bound of speedup by considering the maximum of the longest forward pass and longest backward pass. The speedup bound is given by (\ref{eq:pipedream_speedup_bound}).
\begin{equation}
    \label{eq:pipedream_speedup_bound}
    \sigma_{pipedream} \leq \frac{N_{total}}{\max(\max_l(N_{FP}^l), \max_l(N_{WG}^l + N_{IG}^l))}
\end{equation}
For the values in Table \ref{tab:example_network_cycles}, the longest forward pass takes $13,916$ cycles, and the longest backward pass takes $11,456$ cycles, resulting in $\sigma_{pipedream}$ of $3.37$. Figure \ref{fig:pipedream_map}(a) shows the schedule map for the PipeDream-based scheduling algorithm when $P = 4$. The number of cycles per weight update is $13,916$, resulting in a speedup of $3.37\times$ over the single processor implementation. Figure \ref{fig:pipedream_map}(b) shows the split of the DFG into four subgraphs.

\vspace{-1em}
\subsection{Split backward pass}
\indent The PipeDream-based scheduling algorithm considers the entire backward pass of a layer as a single task. In most cases, this becomes a bottleneck. The backward pass can be split into weight gradient and input gradient tasks. This increases the upper bound of speedup by considering the longest task. The upper bound of speedup is given by (\ref{eq:split_backward_speedup_bound}).
\begin{equation}
    \label{eq:split_backward_speedup_bound}
    \sigma_{split} \leq \frac{N_{total}}{\max(\max_l(N_{FP}^l), \max_l(N_{WG}^l), \max_l(N_{IG}^l))}
\end{equation}
However, for the example network, this does not offer any further speedup as the longest task is the forward pass of the first layer. \vspace{-1em}
\subsection{Fine-grained pipelining}
The speedup of training might not be close to ideal even when the number of processors is less than the upper bound of speedup. In Fig. \ref{fig:layerwise_map}, the speedup is only $1.8\times$ instead of the ideal speedup of $2\times$. This is because the tasks require a different number of clock cycles, which results in some processors being idle. Fine-grained pipelining can be used to balance the pipeline stages such that the number of clock cycles required for each stage is the same, thus reducing the idle time of the processors.

\begin{figure}[htbp]
    \centering
    \vspace{-1em}
    \includegraphics[scale=0.8]{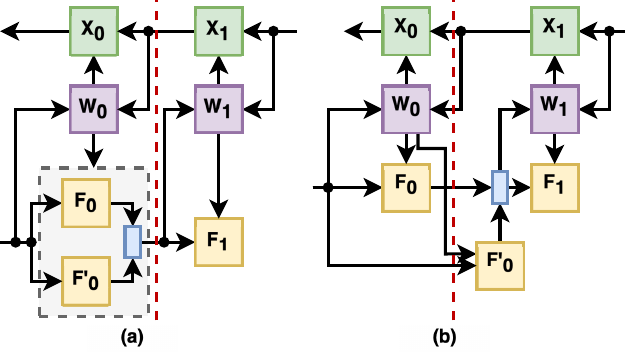} \vspace{-1em}
    \caption{Mapping two layers to two processors by (a) splitting the forward pass in the first layer and (b) moving one part of the forward pass to the next processor.} \vspace{-1em}
    \label{fig:finegrain_forward_move}
\end{figure}

Fine-grained pipelining is done by splitting the tasks into multiple subtasks wherever necessary and assigning them to different processors. The forward pass of a layer needs outputs of the previous layer and weights of the current layer. The output of the forward pass is used as input to the forward pass and weight gradient of the next layer. Therefore, moving a part of the forward pass to the next processor requires more communication of moving part of previous outputs and current weights to the next layer. However, it reduces the cost of moving the outputs of the current forward pass to the next processor. Figure \ref{fig:finegrain_forward_move} shows the process of moving part of the forward pass to the next processor. The task $F_0$ is split into $F_0$ and $F'_0$. The task $F'_0$ is moved to the next processor.

\begin{figure}[htbp]
    \centering
    \vspace{-1em}
    \includegraphics[scale=0.675]{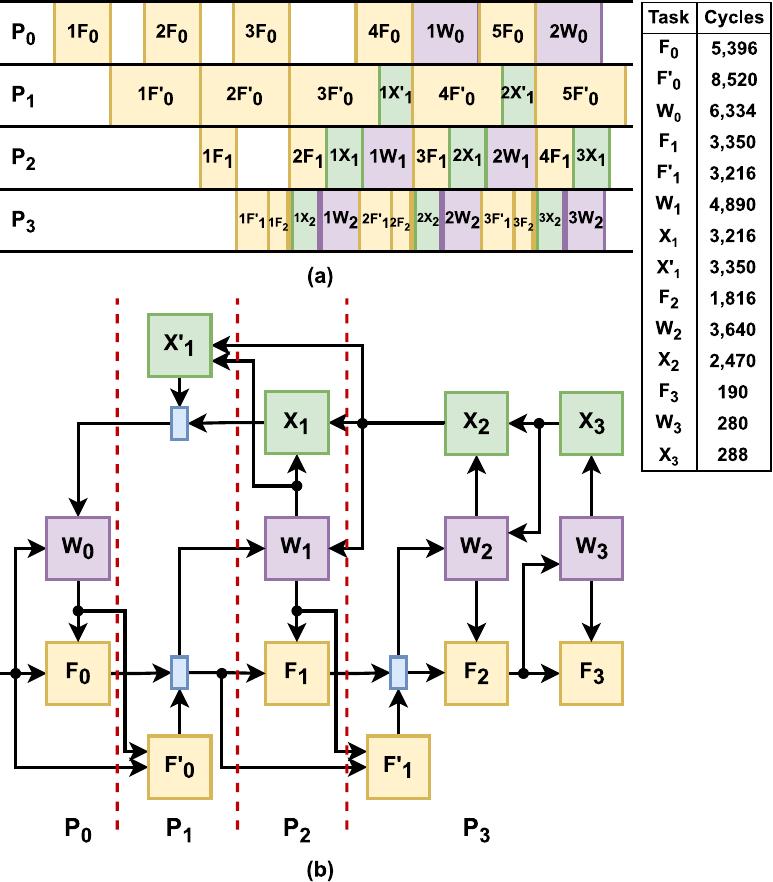} \vspace{-1em}
    \caption{(a) Schedule map for the proposed fine-grained scheduling scheme when $P = 4$ and (b) split of the DFG to four subgraphs.} \vspace{-1em}
    \label{fig:finegrain_map}
\end{figure}

\begin{algorithm}[htbp]
    \caption{First-to-last allocation scheme}
    \label{alg:first_to_last}
    \begin{algorithmic}[1]
        \State \Comment{Compute clock cycles for each task and store in a matrix of size $L \times 3$}
        \State $N_C \gets$ \Call{ComputeCycles}{Network}
        \State \Comment{Allocate tasks to $P$ processors using first to last allocation scheme}
        \State $N_{tot} \gets sum(N_C[l, i])$ for $l \gets 0$ to $L-1, i \gets 0$ to $2$
        \State $N_{ideal} \gets \frac{N_{tot}}{P}$ \Comment{Compute ideal time per processor}
        \State $\alpha \gets 1.1$
        \State flag $\gets$ True
        \While{flag}
            \State $N_{alloc}[p] \gets 0$ for $p \gets 0$ to $P-1$
            \State Layer index, $l \gets 0$
            \State Processor index, $p \gets 0$
            \State Task index, $i \gets 2$
            \While{True}
                \If{$N_{alloc}[p] + N_C[l, i] \leq N_{ideal}$}
                    \State Allocate task $i$ of layer $l$ to processor $p$
                    \State $N_{alloc}[p] \gets N_{alloc}[p] + N_C[l, i]$
                    \If{$i = 0$}
                        \State $l \gets l + 1$
                        \State $i \gets 2$
                    \Else
                        \State $i \gets i - 1$
                    \EndIf
                \Else
                    \If{$i=1$} \Comment{Cannot split weight gradient}
                        \If{$N_C[l,i] / 2 \leq (N_{ideal}-N_{alloc}[p])$}
                            \State Allocate task $i$ of layer $l$ to processor $p$
                            \State $N_{alloc}[p] \gets N_{alloc}[p] + N_C[l, i]$
                            \State $i \gets 0$
                        \EndIf
                    \Else
                        \State $N_{rem} \gets N_{ideal} - N_{alloc}[p]$
                        \State Split task into $N_{rem}$ and $N_C[l, i] - N_{rem}$
                        \State Allocate $N_{rem}$ of task $i$ to processor $p$
                        \State $N_{alloc}[p] \gets N_{alloc}[p] + N_{rem}$
                        \State $N_C[l, i] \gets N_C[l, i] - N_{rem}$
                    \EndIf
                    \State $p \gets p + 1$
                \EndIf
                \If{$p = P-1$}
                    \If{$l = L-1$}
                        \State flag $\gets$ False
                    \Else
                        \State $N_{ideal} \gets \alpha \cdot N_{ideal}$
                    \EndIf
                    \State \textbf{break}
                \EndIf
            \EndWhile
        \EndWhile
        \State \textbf{return} layer\_map, $N_{alloc}$
    \end{algorithmic}
\end{algorithm}

Similarly, the input gradient task can be split into subtasks and moved to the previous processors. Both the forward pass and input gradient splits involve a small overhead of communicating appropriate weights to the respective processor. Splitting the weight gradient task, however, would result in a large overhead of moving either the outputs of the previous layer or the input gradient of the next layer. Therefore, the weight gradient task is not split. Using these conditions, the tasks are allotted to processors using a first-to-last allocation scheme detailed in Algorithm \ref{alg:first_to_last}.

The first-to-last algorithm allocates the tasks in the order of input gradient, weight gradient, and forward pass. The algorithm starts with assigning the first layer's tasks to the first processor and continues allocating the tasks to the next processor until all the processors are full. If all the layers are not assigned, the ideal time constraint is relaxed, and the algorithm is rerun until all the layers are allocated. Similarly, the tasks can be allocated in the order of forward pass, input gradient, and weight gradient from the last layer to the first layer, resulting in a last-to-first allocation scheme.

For a given network, both allocation schemes are run, and the one with the better speedup is chosen. Since the weight gradient task is not split, this becomes the bottleneck for speedup. The maximum speedup for this scheme is given by (\ref{eq:finegrained_speedup_bound}).
\begin{equation}
    \label{eq:finegrained_speedup_bound}
    \sigma_{finegrained} = \frac{N_{total}}{\max_l(N_{WG}^l)}
\end{equation}
For the example network, the longest weight gradient task takes $6,334$ cycles, resulting in a $\sigma_{finegrained}$ of $7.41 \times$, which is more than twice that of the speedup from the PipeDream-based scheduling scheme. Figure \ref{fig:finegrain_map}(a) shows the schedule map for the proposed fine-grained scheduling scheme for $P = 4$. Figure \ref{fig:finegrain_map}(b) shows the modified DFG with the tasks split and moved to execute the schedule given by Fig. \ref{fig:finegrain_map}(a). With this split of the DFG, the number of cycles per weight update is $11,900$, resulting in a speedup of $3.95 \times$, which is very close to the ideal speedup of $4 \times$.
\section{Simulation Results}
\label{sec:results}
\begin{figure}[!h]
    \centering
    \vspace{-1em}
    \includegraphics[scale=0.3]{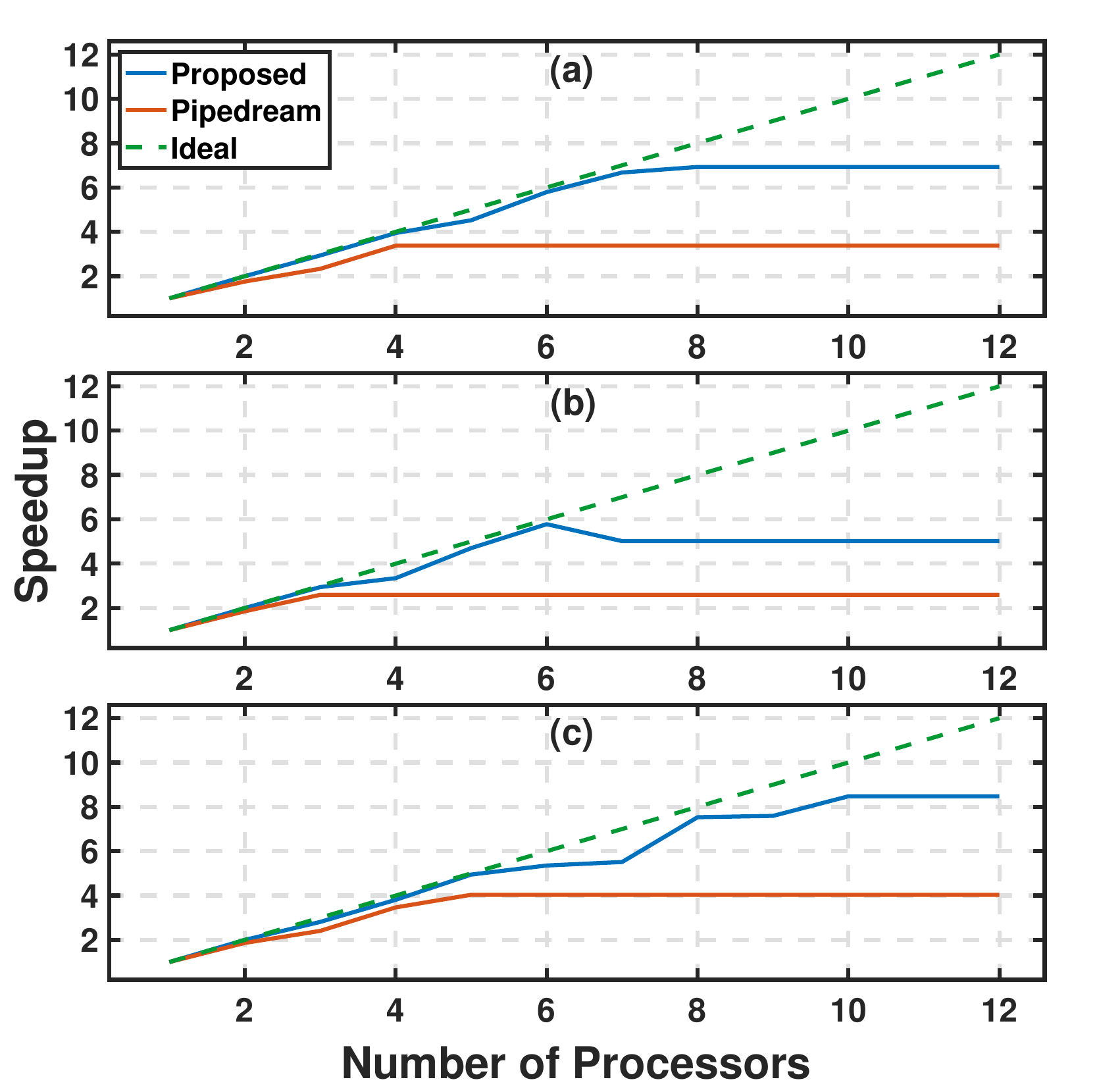} \vspace{-1em}
    \caption{Comparison of the proposed fine-grained scheduling with the PipeDream-based scheduling over single-processor implementation for the (a) MNIST, (b) N-MNIST, and (c) DVS128 Gestures networks. The results are with a batch size of 1 and a $32 \times 32$ systolic array.}
    \label{fig:speedup_plot}
\end{figure}

The proposed fine-grained scheduling scheme is simulated on the networks described in Table \ref{tab:example_network} and compared with the PipeDream-based scheduling scheme for a varying number of processors. Figure \ref{fig:speedup_plot} shows the speedup of both scheduling schemes for the example network with a batch size of 1 when mapped to $32 \times 32$ systolic arrays.

\begin{figure}[!h]
    \centering
    \vspace{-1.5em}
    \includegraphics[scale=0.3]{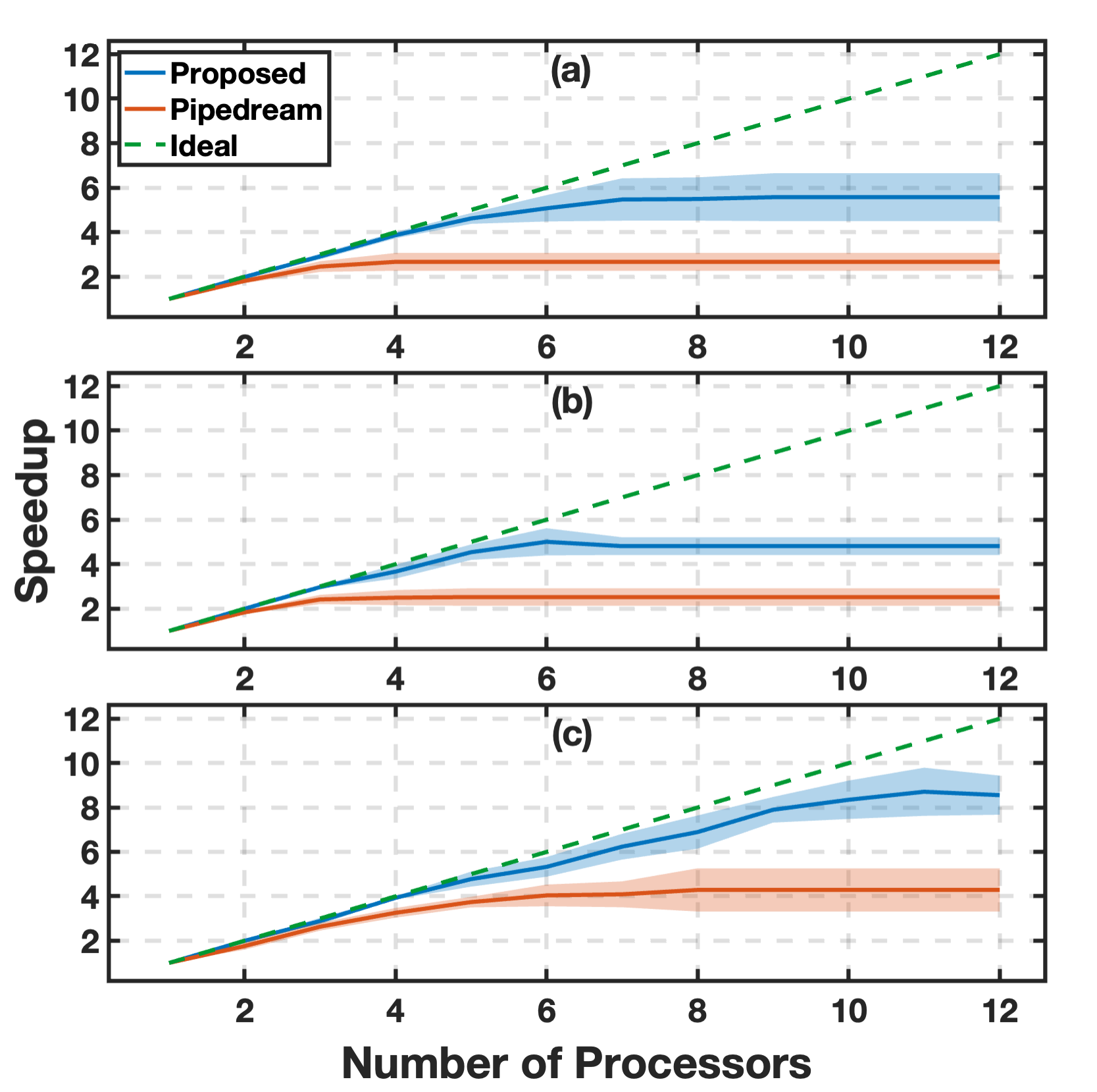} \vspace{-1em}
    \caption{Mean and standard deviation of the speedup with different batch sizes and systolic array sizes for (a) MNIST, (b) N-MNIST, and (c) DVS128 Gestures networks.} \vspace{-0.5em}
    \label{fig:average_speedup}
\end{figure}

To test the effect of batch size and systolic array size on the speedup, both the scheduling algorithms are simulated for different batch sizes and systolic array sizes. Figure \ref{fig:average_speedup} shows the mean and standard deviation of the speedup when batch size is varied from 1 to 128, and the systolic array size is varied from $16 \times 16$ to $256 \times 256$.

\begin{figure}[!h]
    \centering
    \vspace{-1em}
    \includegraphics[scale=0.3]{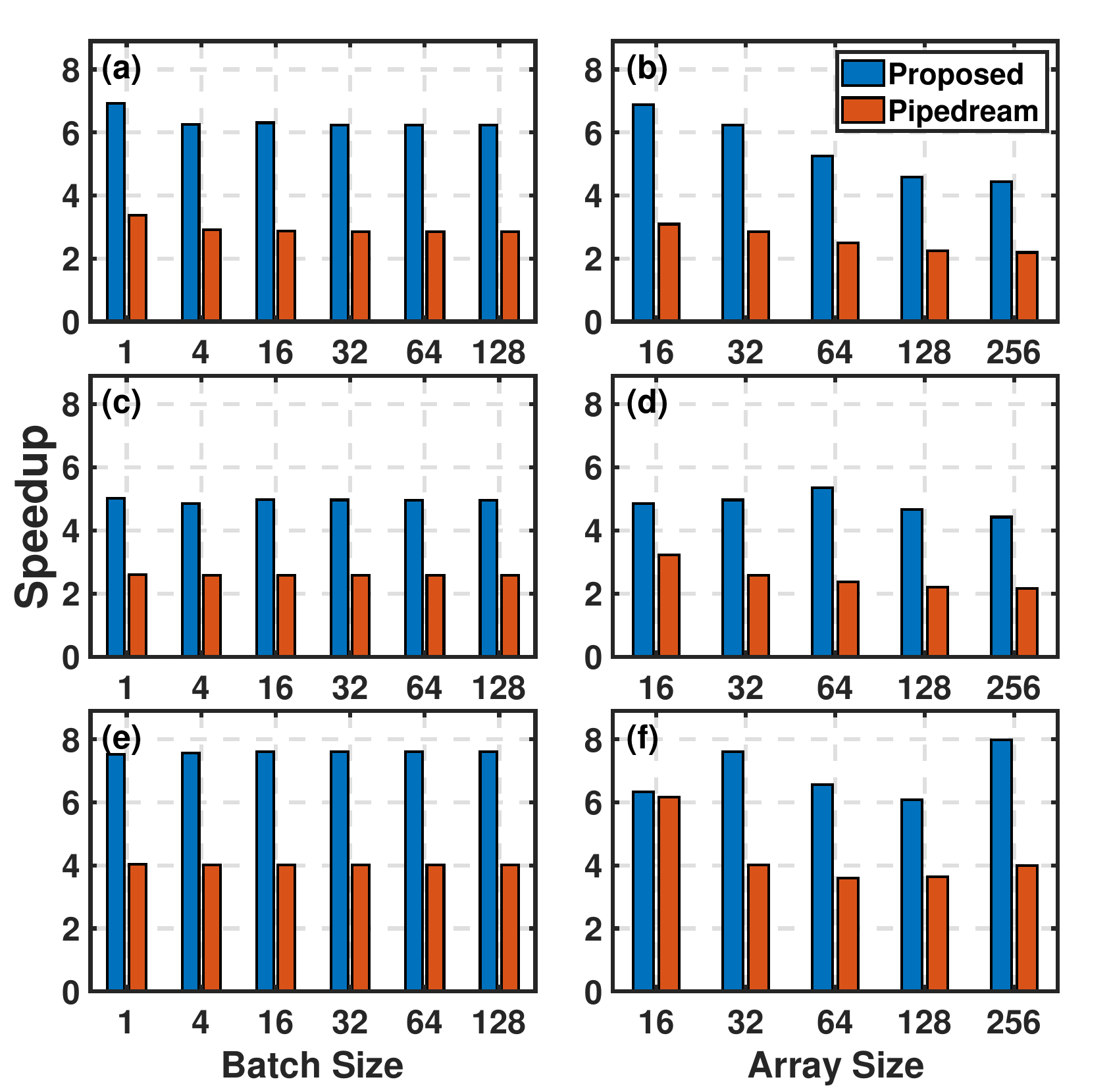} \vspace{-1em}
    \caption{Speedup with 8 processors with (a) varying batch sizes of the MNIST network, (b) varying systolic array sizes of the MNIST network, (c) varying batch sizes of the N-MNIST network, (d) varying systolic array sizes of the N-MNIST network, (e) varying batch sizes of the DVS128 Gestures network, and (f) varying systolic array sizes of the DVS128 Gestures network. Results with varying batch sizes are with a systolic array of size $32 \times 32$, and varying array sizes are with a batch size of 32. The speedup is normalized to one processor.} \vspace{-1em}
    \label{fig:average_batch_array}
\end{figure}

Figure \ref{fig:average_batch_array} shows the individual effect of batch size and systolic array size on the speedup. Batch size does not have a significant impact on the speedup for both the scheduling schemes, as evident in Fig. \ref{fig:average_batch_array}(a), (c), and (e), which show the speedup with varying batch sizes and a $32 \times 32$ systolic array. Figure \ref{fig:average_batch_array}(b), (d), and (f), however, show that the systolic array size affects the speedup of both algorithms. Having smaller systolic arrays is beneficial for the PipeDream algorithm but the proposed algorithm benefits from a few specific array sizes, based on the network structure.

Table \ref{tab:results_summary} summarizes the speedup of both algorithms on the three networks for various processors, averaged over all batch sizes and systolic array sizes. It also shows the additional overhead incurred by the fine-grained scheduling scheme. On average, the proposed fine-grained pipelining and scheduling algorithm achieves a speedup improvement of $65.28\%$ over the PipeDream-based algorithm. The proposed algorithm achieves more than $2\times$ speedup over the PipeDream-based algorithm when using a higher number of processors.

The overhead incurred by the fine-grained scheduling scheme is negligible compared to the total memory requirement of the network. For the network trained on MNIST, the communication required to transfer intermediate outputs and gradients is 3.22 MB on average. The maximum overhead is 16.86 KB, which is $0.51\%$ of the total communication. For the larger network trained on the DVS128 Gestures dataset, the total communication requirement is 982.98 MB on average, while the maximum overhead is $\approx$ 260 KB, which is only $0.03\%$ of the total communication.

The size of the SRAM required in the processors increases for large networks to store the inputs and intermediate computed variables. However, assuming the worst-case scenario of having a single processor and storing all intermediate outputs for the backward pass, the memory requirement is $11.568$ MB, $44.217$ MB, and $2.092$ GB for the MNIST, N-MNIST, and DVS128 Gestures networks, respectively, for a batch size of $32$. This requirement reduces almost linearly with an increase in number of processors.
\begin{table}[htbp]
    \renewcommand{\arraystretch}{1.1}
    \centering
    \caption{Summary of results for the two scheduling schemes. Results averaged over all batch sizes and systolic array sizes.}
    \label{tab:results_summary}
    \begin{tabular}{|c|c|c|c|c|c|}
        \hline
        \hline
        \multirow{2}{*}{\textbf{Network}} & \multirow{2}{*}{\Centerstack{\textbf{No. of} \textbf{Procs.}}} & \multicolumn{3}{c|}{\textbf{Speedup}} & \multirow{2}{*}{\Centerstack{\textbf{Overhead} \textbf{(KB)}}} \\ \cline{3-5}
        & & \textbf{PipeDream} & \Centerstack{\textbf{Fine} \textbf{Grain}} & \Centerstack{\textbf{Improv.} \textbf{(\%)}} & \\
        \hline
        \hline
        \multirow{7}{*}{\textbf{MNIST}} & 1 & 1.00 & 1.00 & 0.00 & 0.00 \\
        \cline{2-6}
        & 2 & 1.81 & 1.97 & 8.94 & 0.73 \\
        \cline{2-6}
        & 4 & 2.67 & 3.78 & 41.84 & 2.62 \\
        \cline{2-6}
        & 6 & 2.67 & 5.04 & 88.84 & 9.83 \\
        \cline{2-6}
        & 8 & 2.67 & 5.49 & 105.86 & 10.40 \\
        \cline{2-6}
        & 10 & 2.67 & 5.57 & 108.92 & 16.86 \\
        \cline{2-6}
        & 12 & 2.67 & 5.57 & 108.92 & 16.86 \\
        \hline
        \hline
        \multirow{7}{*}{\textbf{N-MNIST}} & 1 & 1.00 & 1.00 & 0.00 & 0.00 \\
        \cline{2-6}
        & 2 & 1.83 & 2.00 & 8.92 & 0.56 \\
        \cline{2-6}
        & 4 & 2.49 & 3.67 & 50.67 & 2.08 \\
        \cline{2-6}
        & 6 & 2.52 & 5.00 & 101.55 & 4.03 \\
        \cline{2-6}
        & 8 & 2.52 & 4.81 & 94.02 & 3.92 \\
        \cline{2-6}
        & 10 & 2.52 & 4.81 & 94.02 & 3.92 \\
        \cline{2-6}
        & 12 & 2.52 & 4.81 & 94.02 & 3.92 \\
        \hline
        \hline
        \multirow{9}{*}{\Centerstack{\textbf{DVS128} \textbf{Gestures}}} & 1 & 1.00 & 1.00 & 0.00 & 0.00 \\
        \cline{2-6}
        & 2 & 1.76 & 1.99 & 14.31 & 9.00 \\
        \cline{2-6}
        & 4 & 3.26 & 3.93 & 21.35 & 54.60 \\
        \cline{2-6}
        & 6 & 4.04 & 5.33 & 33.70 & 96.64 \\
        \cline{2-6}
        & 8 & 4.29 & 6.89 & 67.60 & 133.95 \\
        \cline{2-6}
        & 10 & 4.29 & 8.35 & 102.75 & 194.10 \\
        \cline{2-6}
        & 12 & 4.29 & 8.56 & 106.39 & 260.18 \\
        \cline{2-6}
        & 14 & 4.29 & 9.04 & 113.98 & 252.56 \\
        \cline{2-6}
        & 16 & 4.29 & 9.87 & 134.73 & 254.70 \\
        \hline
        \hline
    \end{tabular}
\end{table}

\section{Conclusion}

This paper discusses the general modeling of Spiking Neural Networks based on the Leaky Integrate-and-Fire model, and their training using the backpropagation algorithm. The dataflow graph of training is then pipelined and retimed using delayed gradients in an attempt to map it to multiple processors. The design of typical systolic array-based processors is discussed, along with their modeling to estimate clock cycles necessary for executing various tasks. Using the estimated clock cycles, the dataflow graph is split in various ways using pre-existing algorithms. A fine-grained pipelining and scheduling scheme is then proposed to improve the throughput of training over conventional methods. The proposed scheme is evaluated on three networks, and the results show an average of $\approx 65\%$ improvement in throughput with upward of $> 100\%$ improvement in some cases, with a small drop in accuracy for larger networks. The overhead incurred by the proposed scheme is $\leq 0.5\%$ compared to the total communication requirement of the network. The proposed scheme assumes that the neural networks have only convolutional or fully-connected layers. The future scope of this work includes extending the proposed scheme to networks with other types of layers, such as normalization and residual layers, which can improve the accuracies of deeper networks. Further, a hybrid multiprocessor training and single processor fine-tuning approach can be explored to improve the accuracy of the networks trained using the proposed scheme.

\begin{IEEEbiography}[{\includegraphics[width=1in,height=1.25in,clip,keepaspectratio]{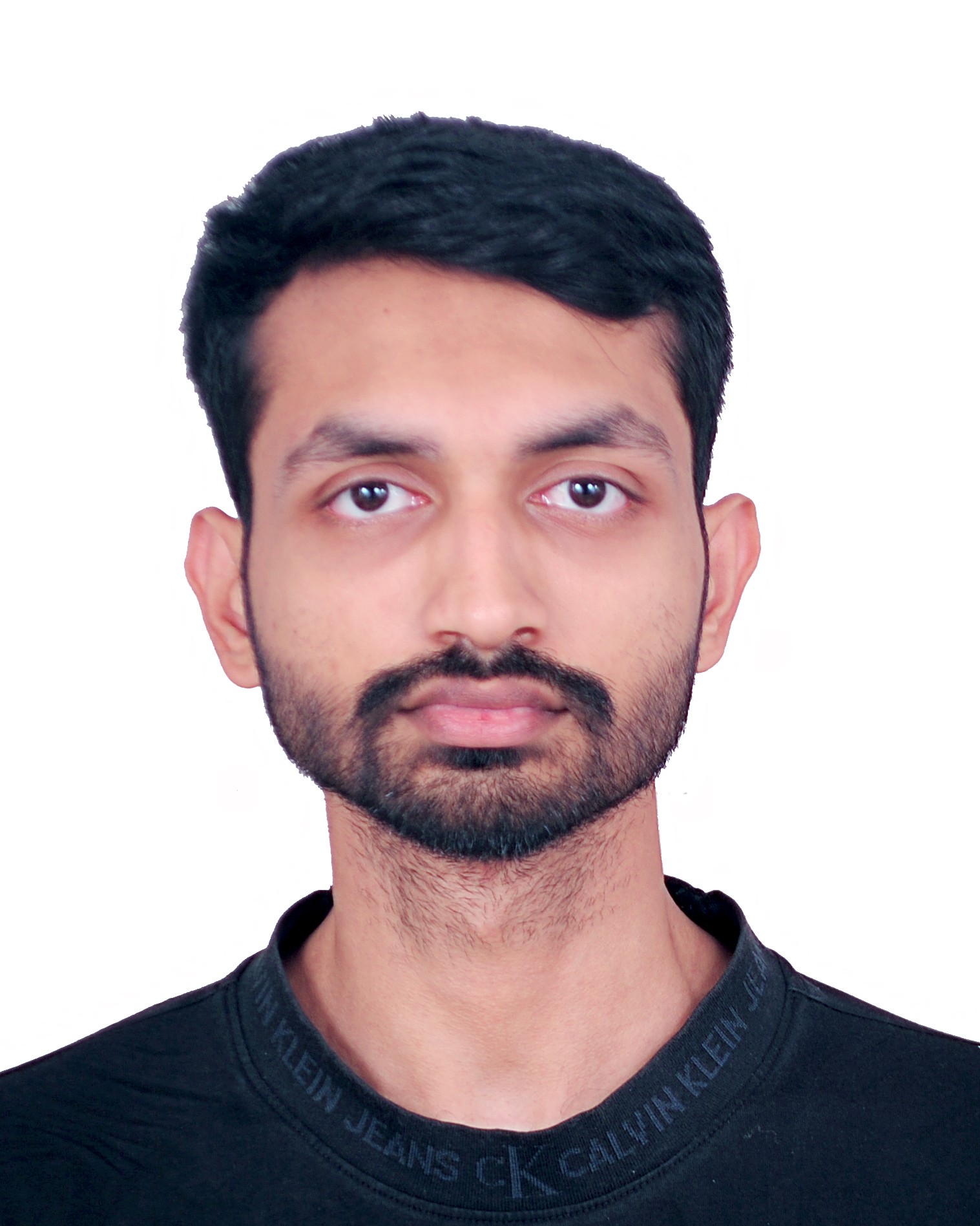}}]{Sai Sanjeet} received the Dual Degree (B. Tech. and M. Tech.) in Electronics and Electrical Communication Engineering from the Indian Institute of Technology Kharagpur, Kharagpur, India, in 2021 with specialization in the area of Visual Image Processing and Embedded Systems. He is currently pursuing his PhD at the University at Buffalo, State University of New York, Buffalo, in the area of analog and mixed-signal hardware accelerators for machine learning applications. His research interests include machine learning, neuro-inspired computing, analog and mixed-signal circuit design, and signal processing.
He was part of a collaborative group from the University of Tokyo which received 3rd place in the 30th Intl. Workshop on Logic and Synthesis (IWLS) Design Contest 2021. He is also a recipient of the National Talent Search Examination (NTSE) fellowship and the Kishore Vaigyanik Protsahan Yojana (KVPY) fellowship.

\end{IEEEbiography}

\begin{IEEEbiography}[{\includegraphics[width=1in,height=1.25in,clip,keepaspectratio]{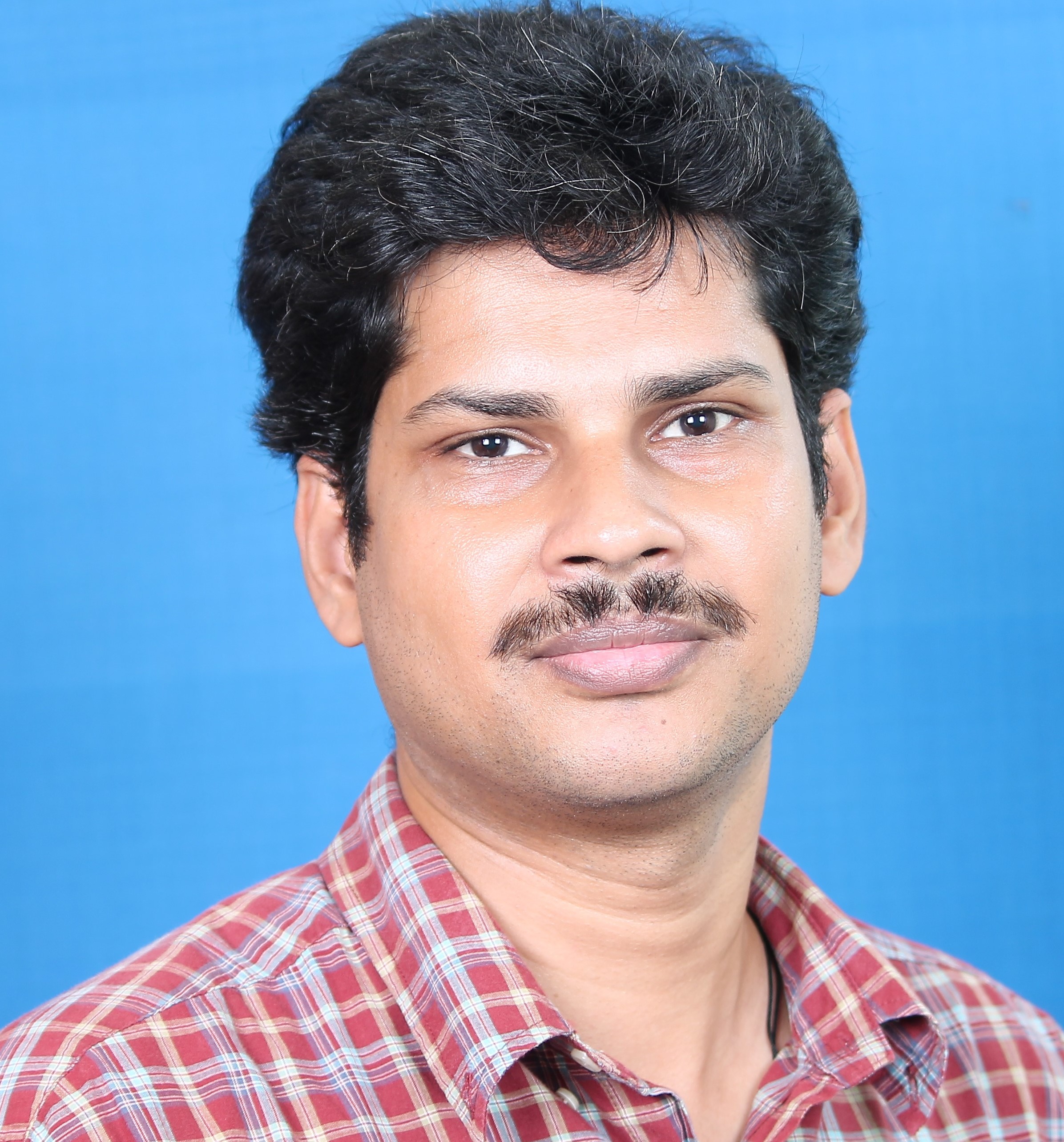}}]{Bibhu Datta Sahoo} received the B.Tech. degree in electrical engineering from the Indian Institute of Technology Kharagpur, Kharagpur, India, in 1998, the M.S.E.E. degree from the University of Minnesota, Minneapolis, MN, USA, in 2000, and the Ph.D.E.E. degree from the University of California, Los Angeles, CA, USA, in 2009. From 2000 to 2006, he was with Broadcom Corporation, Irvine, CA, USA,  From December 2008 to February 2010, he was with Maxlinear Inc., Carlsbad, CA, USA, where he was involved in designing integrated circuits for CMOS TV tuners. From March 2010 to November 2010, he was a Post-Doctoral Researcher with the University of California, Los Angeles, CA, USA. From December 2011 to April 2015, he was an Associate Professor with the Department of Electronics and Communication Engineering, Amrita University, Amritapuri, India. From January 2016 to April 2017 he was on sabbattical from Amrita University and was a Research Scientist at University of Illinois at Urbana-Champaign. From August 2017 to August 2023 he has been Associate Professor in the Department of Electronics and Electrical Communication Engineering at Indian Institute of Technology Kharagpur, Kharagpur, India. Since September 2023 he has been a Professor in the Department of Electrical Engineering at University at Buffalo, State University of New York. His research interests include data converters, signal processing, and analog and mixed signal circuit design.
He received the 2008 Analog Devices Outstanding Student Designer Award and was the co-recipient of the 2013 CICC Best Paper Award. He was the Associate Editor of IEEE Transactions on Circuits and Systems-II from August 2014 to December 2015. Since Dec. 2018 he has been the Associate Editor of IEEE Open Journal of Circuits and Systems.

\end{IEEEbiography}

\begin{IEEEbiography}[{\includegraphics[width=1in,height=1.25in,clip,keepaspectratio]{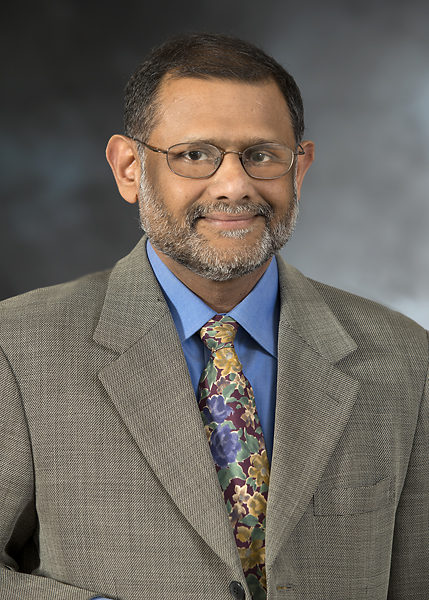}}]{Keshab K. Parhi} (Fellow, IEEE) received the B.Tech. degree from the Indian Institute of Technology (IIT), Kharagpur, in 1982, the M.S.E.E. degree from the University of Pennsylvania, Philadelphia, in 1984, and the Ph.D. degree from the University of California, Berkeley, in 1988. He has been with the University of Minnesota, Minneapolis, since 1988, where he is currently the Erwin A. Kelen Chair and a Distinguished McKnight University Professor in the Department of Electrical and Computer Engineering. He has published over 700 papers, is the inventor of 36 patents, and has authored the textbook VLSI Digital Signal Processing Systems (Wiley, 1999). His current research addresses VLSI architecture design of machine learning and signal processing systems, hardware security, and data-driven neuroengineering and neuroscience. Dr. Parhi is the recipient of numerous awards including the 2003 EEE Kiyo Tomiyasu Technical Field Award; and the 2017 Mac Van Valkenburg award, the 2012 Charles A. Desoer Technical Achievement award, and a Golden Jubilee medal in 1999, from the IEEE Circuits and Systems Society. He served as the Editor-in-Chief of the IEEE Trans. Circuits and Systems, Part-I during 2004 and 2005, and currently serves as the Editor-in-Chief of the IEEE Circuits and Systems Magazine. He is a Fellow of the American Association for the Advancement of Science (AAAS), the Association for Computing Machinery (ACM), the American Institute of Medical and Biological Engineering (AIMBE), and the National Academy of Inventors (NAI).
\end{IEEEbiography}

\end{document}